\documentclass[preprint,aps ,nofootinbib]{revtex4}
\usepackage{graphicx}
\pdfoutput=1
\usepackage{epsfig}
\usepackage{amsmath}
\usepackage{amsfonts}
\usepackage{amssymb}
\usepackage{color}
\usepackage{dcolumn}
\usepackage{hyperref}
\usepackage{multirow}
\usepackage{multirow,gensymb}
\usepackage{booktabs}
\usepackage{tikz}
\usetikzlibrary{calc}

\providecommand{\U}[1]{\protect\rule{.1in}{.1in}}

\newcommand{\f}{\begin{equation}}
\newcommand{\ff}{\end{equation}}
\newcommand{\fa}{\begin{eqnarray}}
\newcommand{\ffa}{\end{eqnarray}}

\begin{document}
\title{The commensurate state and lock-in in a holographic model}
\author{Yi Ling $^{1,2}$}
\email{lingy@ihep.ac.cn}
\author{Peng Liu $^{3}$}
\email{phylp@email.jnu.edu.cn}
\author{Meng-He Wu$^{4}$}
\email{mhwu@njtc.edu.cn} \affiliation{$^1$
  Institute of High Energy
  Physics, Chinese Academy of Sciences, Beijing 100049, China\\ $^2$
  School of Physics, University of Chinese Academy of Sciences,
  Beijing 100049, China \\ $^3$ Department of Physics and Siyuan Laboratory,
  Jinan University, Guangzhou 510632, China\\
  $^4$  College of Physics and Electronic Information Engineering, Neijiang Normal University, Neijiang 641112, China. }

\begin{abstract}

  We study a holographic model in which the striped structure of charge density is spontaneously formed over an ionic lattice which breaks the translational symmetry explicitly. The effect of commensurate lock-in between the spontaneous stripes and the ionic lattice is observed when the lattice amplitude is large enough. We investigate the optical conductivity as a function of frequency in commensurate state and compare its characteristics during the phase transition from metallic phase to insulating phase. Notably, we find that the DC resistivity in lock-in state increases algebraically with lowering temperature, which is in line with the phenomenon observed in the holographic model for simulating the experimental behavior of Mott insulator in \cite{Andrade:2017ghg}. In addition, at lower temperature the pinning effect is observed for both unlock-in and lock-in states. This holographic model successfully demonstrates the commensurate lock-in signatures, and provides more information for understanding the interplay between ionic lattices  and electronic lattices by holography.

\end{abstract}

\maketitle

\section{Introduction}

Gauge/Gravity duality as a powerful tool for providing a comprehensive understanding on the phase diagram of high temperature superconductivity has made substantial progress since the seminal work in \cite{Gubser:2008px,Hartnoll:2008vx,Hartnoll:2008kx}. Firstly, to guarantee that the divergent behavior of the imaginary part of optical conductivity at zero frequency results from the spontaneous breaking of $U(1)$ symmetry rather than the translational invariance owned by the original model, the lattice structure has been introduced into the holographic system to relax the momentum such that the standard Drude behavior of the conductivity is observed in metallic phase of the dual system\cite{Horowitz:2012ky,Horowitz:2012gs,Horowitz:2013jaa,Vegh:2013sk,Ling:2013nxa,Blake:2013owa,Donos:2013eha,Ling:2014laa,Ling:2017naw,Donos:2014yya,Liu:2012tr,Ling:2016lis,Andrade:2013gsa,Wu:2018zdc,Ling:2014bda,Baggioli:2014roa,Donos:2014uba,Baggioli:2021xuv,Donos:2012js,Wang:2021jfu}. Secondly, to describe the novel metal-insulator transition by charge density waves, which is an ordered state with the modulation of electronic density in a periodic structure, the mechanism leading to the spontaneous breaking of the translational symmetry has also been extensively studied in holographic literature and two prominent features of CDW, namely the pseudogap and pinning effects, have been revealed \cite{Baggioli:2019abx,Baggioli:2020edn,Ooguri:2010xs,Donos:2011bh,Donos:2013wia,Donos:2013gda,Withers:2013loa,Withers:2013kva,Ling:2014saa,Andrade:2017cnc,Andrade:2020hpu,Ling:2020qdd,Baggioli:2022pyb,Amoretti:2021fch,Li:2018vrz,Song:2019rnf,Alberte:2017oqx,Amoretti:2017axe,Ammon:2019wci}. Thirdly, recently great efforts have been made to construct the holographic models simulating the behavior of Mott insulator which is also an essential step towards the comprehensive understanding of the phase diagram for high temperature superconductivity \cite{Ling:2015epa,Baggioli:2016oju,Andrade:2017ghg,Kiritsis:2015hoa,Cremonini:2016rbd,Baggioli:2015dwa,Baggioli:2015zoa,Ling:2015exa,Cai:2020nyd,Edalati:2010ww,Edalati:2010ge,Fujita:2014mqa}.

One of the key ingredients to simulate the Mott insulator is its commensurate feature, which states that the period of the ionic lattice has a rational relationship with the period of the charge density waves, or the electronic lattice. Namely, the ratio of these two periods (or the ratio of two wave-vectors) is a ratio of two integers. In general, the ionic lattice, introduced to relax the momentum, may have its wave-vector manually set. Conversely, the electronic lattice, which arises from the spontaneous breaking of translational symmetry, has its wave-vector determined by the instability of system. Therefore, these two periodic structures are fundamentally independent, and their wave-vectors' ratio can be an arbitrary number, leading to either a commensurate state (rational ratio) or an incommensurate state (irrational ratio). On the other hand, the ionic lattice can provide the periodic background potential for CDW. As such, the lattice structure inevitably impacts the CDW pattern and stability. Specifically, a strong lattice can force the spontaneous formation of CDW stripes to align commensurately with the underlying lattice. This commensurate ``lock-in effect" reveals the intimate relationship between the emergent CDW order and the underlying lattice. It is a significant phenomenon which results in profound consequences such as the formation of electronic density gaps.

In holographic approach, the above picture has been clearly demonstrated with toy models and cartoons in \cite{Andrade:2017leb} and \cite{Krikun:2017cyw}. By constructing a specific holographic model, it is found in \cite{Andrade:2017leb} that with the increase of the lattice amplitude, the wave-vector of CDW deviates from its value in the absence of lattice, and finally coincides with multiples of the lattice wave-vector, which is the first observation of the commensurate lock-in in holographic literature.

So far, the lattices in holographic models may be classified into two general classes based on their distinct structure, namely the homogeneous lattices and inhomogeneous lattices.
Homogeneous lattices, such as axion fields and Q-lattices, break translational symmetry but may not exhibit an explicit periodic structure along the spatial direction. In such cases, it has been found that the commensurate lock-in effect cannot be observed even with the presence of pinning effects on the CDW
\cite{Andrade:2015iyf, Jokela:2017ltu,Andrade:2018gqk,Ling:2020mwm}. It was later realized that achieving the commensurate lock-in effect in holography requires an inhomogeneous lattice that explicitly displays a periodic structure \cite{Andrade:2017leb}.

Incorporating an inhomogeneous lattice background brings technical challenges, as it involves in solving nonlinear partial differential equations to obtain the lattice background and to analyze its instability, which signals the production of the CDW.
Previously a holographic model with a topological term has been applied to attack this problem and some insightful results have been obtained for the holographic construction of Mott insulators \cite{Andrade:2017ghg}. As far as we know, currently all the knowledge on the commensurate lock-in and its relevant phenomenon in holographic framework is heavily based on this single model \cite{Andrade:2017leb, Jokela:2017ltu, Krikun:2017cyw, Andrade:2018gqk}.
However, it is believed that the commensurate state and lock-in effects should be general phenomena and should be observed in other holographic models with inhomogeneous lattices.
Furthermore, the topological term introduced in this model is not an essential ingredient and many alternative mechanisms could lead to the instability of the background. More importantly, due to the effect of this topological term, the magnetic field must emerge due to the instability of the background such that current density waves have to be incorporated with the emergence of CDW \cite{Donos:2011bh,Andrade:2017ghg}. As pointed out in the original papers on this model \cite {Donos:2011bh,Donos:2013wia,Donos:2013gda}, the appearance of current density waves breaks  parity (P) and time-reversal (T) invariance, which of course are not necessary features of CDW.

Our goal in this paper is to provide an affirmative answer to the question whether the lock-in effects can be implemented in other holographic models with inhomogeneous lattices.
Specifically, we explore the commensurate states and lock-in effect based on a simple holographic model that excludes the need for a topological term, resulting in the absence of current density waves in the background and preserving PT symmetry \cite{Donos:2011bh,Donos:2013gda,Ling:2014saa}.
Furthermore, we note a gap in the current literature regarding the comparison between the optical conductivity of the lock-in state and the unlock-in state. To address this, we compute the optical conductivity for both states separately. This approach aids in revealing the features of the commensurate lock-in states, thereby illuminating the use of holographic models to understand fundamental problems in strongly coupled systems.

The paper is organized as follows. First, we introduce the setup of holographic model with two U(1) gauge fields in Sec. \ref{sec:setup}. In Sec. \ref{sec:instability}, we study the instability of the background with ionic lattices by linear perturbations, and then obtain the phase diagram for the production of CDW. Furthermore, we obtained various phase diagrams by changing the amplitude of the lattice, and the commensurate lock-in effect is demonstrated.  In Sec. \ref{sec:numerical}, we present the numerical result for the background with both ionic lattice and CDW, and the constituents of charge density after the phase transition is analyzed. In Sec. \ref{sec:optical}, we compute the optical conductivity of the gauge field $A$  in commensurate state and compare its characteristics during the phase transition from metallic phase to insulating phase, with the focus on the difference of its behavior between the lock-in state and unlock-in state. Our conclusions and discussions are presented  in Sec. \ref{sec:dis}.

\section{The holographic setup}\label{sec:setup}
The early investigation on the striped phase of CDW in the context of holographic gravity can be found in \cite{Donos:2011bh,Donos:2013wia,Withers:2013loa,Withers:2013kva,Donos:2013gda,Ling:2014saa,Withers:2014sja,Cai:2017qdz,Ooguri:2010xs,Rozali:2013ama,Cremonini:2018xgj}, where several models were proposed to induce the instability of the background. To avoid the involvement of current density waves, in this paper we start with a model with two gauge fields, and the action is given by \cite{Donos:2013gda,Ling:2014saa},
\begin{equation}\label{eq:action}
  S=\int d^{4} x \sqrt{-g}\left(R-\frac{1}{2}(\partial \psi)^{2} -V(\psi)-\frac{Z_A(\psi)}{4} F^{2}-\frac{Z_B(\psi)}{4} G^{2}- \frac{Z_{AB}(\psi)}{2} F G \right),
\end{equation}
where $\psi$ denotes a scalar dilaton field,
while $F=dA$ and $G=dB$ are field strength tensors, with $A$ and $B$ being two $U(1)$ gauge fields. The coupling between gauge fields and the scalar field is specified by $Z_{A}(\psi)=1-\frac{\beta}{2} L^{2} \psi^{2}$, $Z_B(\psi)=1$, and $Z_{A B}(\psi)=\frac{\gamma}{\sqrt{2}} L \psi$, where $L$ is the AdS radius, and $\beta$ and $\gamma$ are coupling constants. For concreteness, we set $\beta=-450$ and $\gamma=37.3$ throughout this paper\footnote{We remark that other values of $\beta$ and $\gamma$ will render qualitatively similar results as shown in this paper.}. The scalar field potential $V(\psi)$ is given by $V(\psi)=\frac{1}{2} m^{2} \psi^{2}-\frac{1}{L^2}$, with the mass $m^2=-2/l^2$. For simplicity, we set the AdS radius $L^2=1/24$ and $l^2=1/4$.

The equations of motion can be derived from Eq. \eqref{eq:action} as follows,
\begin{equation}\label{eq:eom}
  \begin{aligned}
    R_{\mu \nu}-T_{\mu \nu}^{\psi}-T_{\mu \nu}^{A} -T_{\mu \nu}^{B}-T_{\mu \nu}^{AB}                                             & =0, \\
    \nabla^{2} \psi-\frac{1}{4} Z_{A}^{\prime} F^{2} -\frac{1}{4} Z_{B}^{\prime} G^{2}-\frac{1}{2} Z_{AB}^{\prime} FG-V^{\prime} & =0, \\
    \nabla_\mu\left(Z_A F^{\mu \nu}+Z_{A B} G^{\mu \nu}\right)                                                                   & =0, \\
    \nabla_\mu\left(Z_B G^{\mu \nu}+Z_{A B} F^{\mu \nu}\right)                                                                   & =0,
  \end{aligned}
\end{equation}
where $R_{\mu \nu}$ is Ricci curvature tensor, while the prime denotes the derivative with respect to $\psi$. The stress-energy tensors for the scalar field $\psi$, the gauge fields $A$, $B$ and their interaction are represented by $T_{\mu \nu}^{\psi}$, $T_{\mu \nu}^{A}$, $T_{\mu \nu}^{B}$ and $T_{\mu \nu}^{A B}$ respectively, as shown below,
\begin{equation}
  \begin{aligned}
    T_{\mu \nu}^{\psi} & =\frac{1}{2} \nabla_{\mu} \psi \nabla_{\nu} \psi+\frac{1}{2} V g_{\mu \nu},               \\
    T_{\mu \nu}^{A}    & =\frac{Z_{A}}{2}\left(F_{\mu \rho} F_{\nu}{^{\rho}}-\frac{1}{4} g_{\mu \nu} F^{2}\right), \\
    T_{\mu \nu}^{B}    & =\frac{Z_{B}}{2}\left(G_{\mu \rho} G_{\nu}{^{\rho}}-\frac{1}{4} g_{\mu \nu} G^{2}\right), \\
    T_{\mu \nu}^{A B}  & =Z_{A B}\left(F_{(\mu|\rho|} G_{\nu)}^\rho-\frac{1}{4} g_{\mu \nu} F G\right).
  \end{aligned}
\end{equation}

The planar AdS-Reissner-N\"ordstrom (AdS-RN) black brane is the simplest charged solution of Eq. \eqref{eq:eom},
\begin{equation}
  \begin{aligned}
    d s^{2}= & \frac{1}{z^{2}}\left[-(1-z) p(z)  d t^{2}+\frac{ d z^{2}}{(1-z) p(z)}+d x^{2}+d y^{2}\right], \\
    A=       & \mu(1-z)  d t,  \quad B=0 , \quad \psi =0,
  \end{aligned}
\end{equation}
where $p(z)=4\left(1+z+z^{2}-\frac{\mu^{2} z^{3}}{16}\right)$, and $\mu$ is the chemical potential of the dual field theory and will be treated as the unit of the dual system. Obviously, this static spacetime possesses translation symmetry along the planar directions $x$ and $y$. The Hawking temperature of the black brane is given by,
\begin{equation}\label{eq_tem}
  T /\mu=\left(48-\mu^{2}\right) /\left(16 \pi \mu\right).
\end{equation}

Next, we introduce the lattice structure to explicitly break the translation symmetry and then investigate its instability under perturbations. As stressed in the introduction, to observe the lock-in effect, we will consider the inhomogeneous lattice which explicitly exhibits the periodic dependence of the spatial coordinate. Without loss of generality, we consider an ionic lattice by setting the chemical potential along the $x$ direction as
\begin{equation}\label{eq_il}
  \mu(x)=\mu(1+ \lambda \cos kx),
\end{equation}
where $\lambda$ and $k$ are the lattice amplitude and the wave-vector, respectively.
To describe  the spacetime background with inhomogeneous lattices, we adopt the following ansatz,
\begin{equation}
  \begin{aligned}\label{latticebg}
    d s^{2} & =\frac{1}{z^{2}}\left[-Q_{t t}(1-z) p(z) d t^{2}+Q_{z z} \frac{d z^{2}}{(1-z) p(z) }+Q_{x x}\left(d x+z^{2}Q_{z x} d z\right)^{2}+Q_{y y}d y^{2} \right], \\
    A=      & \mu(x)(1-z)  a d t, \quad B=0 , \quad \psi =0,
  \end{aligned}
\end{equation}
where $Q_{t t},Q_{z z},Q_{x x},Q_{z x},Q_{y y}$ and $a$ are functions of $x$ and $z$, and should be obtained by solving the equations of motion, which are a group of partial differential equations. Obviously, setting $Q_{t t}=Q_{z z}=Q_{x x}=Q_{y y}=a=1$ and $Q_{z x}=\lambda=0$ recovers the AdS-RN black brane solution. We apply the DeTurck method to solve these equations\cite{Headrick:2009pv}. In addition, we remark that with the regular conditions at horizon, the DeTurck term leads to a boundary condition $Q_{t t}(1,x)=Q_{z z}(1,x)$, implying that these new black brane solutions with spatial modulation have the same temperature as planar AdS-RN black branes, given by Eq. \eqref{eq_tem}.

\section{The instability of the background with ionic lattice }\label{sec:instability}

In this section, we focus on the interaction between the ionic lattice and the CDW at the linear perturbation level, and then demonstrate the commensurate lock-in effect by plotting the variation of the phase diagram with the lattice amplitude.
For this purpose, one may either start with the background with ionic lattice and analyze its instability to generate CDW, or start with the background with  CDW and subsequently introduce the ionic lattice to observe how it impacts CDW. We choose the former approach, as the involvement of the ionic lattice alters the critical temperature and the critical wave-vector of CDW simultaneously. In particular, with the increase of the lattice amplitude, these changes become dramatic. Thus, we will start with a background of ionic lattices and then perturb the system
to analyze the linearized equations of motion. By treating these equations as a generalized eigenvalue problem, we diagnose the stability of the perturbations by identifying the minimum real eigenvalues.

\begin{figure} [h]
  \center{
    \includegraphics[width=0.5\textwidth]{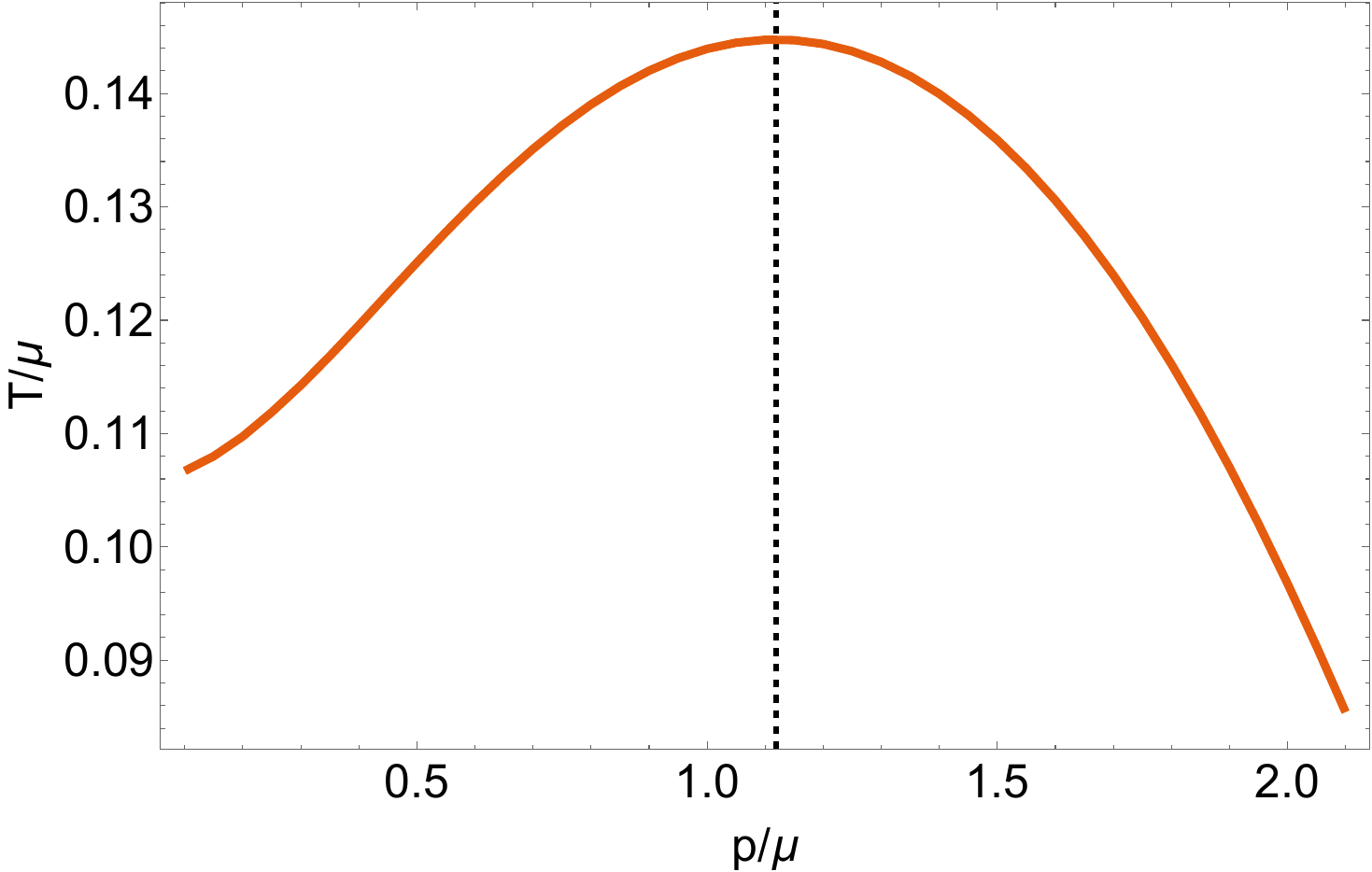}\ \hspace{0.05cm}
    \caption{\label{fig1} The instability region of the background with translation symmetry.}}
\end{figure}

As the first step, we will consider the instability of standard AdS-RN black brane.
We remark that in this model the possible instability is introduced by the coupling term $Z_{A}(\psi)$ rather than the coupling term $Z_{AB}(\psi)$\cite{Ling:2020qdd}, and the detailed analysis based on the extremal black hole at zero temperature limit can be found in \cite{Donos:2013gda}. The key point is that the near horizon geometry of the extremal black hole is $AdS_2$, which is unstable under the perturbations beyond the $BF$ bound. Specifically, we turn on the second gauge field $B$ as well as the scalar field $\psi$, and consider sinusoidal perturbations with wavenumber $ p $ along $ x $ direction with the form as described by,
\begin{equation}\label{eq_per1}
  \begin{aligned}
    \delta \psi & =\delta \psi^{p}(z)\cos(p x), \\
    \delta B    & =\delta B^{p}(z)\cos(p x),
  \end{aligned}
\end{equation}
which preserves the existing $z$-dependence of the background solution. Here $p$ is a real number.
Plugging Eq. \eqref{eq_per1} into the equations of motion about $B$ and $\psi$, and the resultant linear equations of perturbation can be reformulated as a generalized eigenvalue problem about $p$,
\begin{equation}\label{eq:o1}
  \mathcal{O}^{{p}}(z)
  \left(\begin{array}{l}
      \delta \psi^{{p}} \\
      \delta B^{{p}}
    \end{array}\right)=0.
\end{equation}
One can find the eigenvalues of the operator $\mathcal{O}^{{p}}(z)$ by discretizing  Eq. \eqref{eq:o1} and solving the resulting matrix equation.
It turns out that nontrivial solutions can be found when the Hawking temperature is dropped down to some critical value, signaling the instability of the background. Finding the minimum real eigenvalues of $p$ at a given temperature allows us to map out the unstable region in the $ p-T $ plane, as illustrated in Fig. \ref{fig1}. In this figure the dome below the curve is the unstable region under linear perturbations, and one may read out the critical temperature $T_c/\mu  \approx 0.145$ and the critical wavenumber $p_c / \mu \approx 1.1$ by locating the tip of the dome, which is marked by a dashed line in the figure. The existence of such unstable modes implies that new solutions with spatial modulations along $x$ can be found for the background by
solving all the equations of motions in a non-perturbative manner.

Next, we
investigate the instability of the background with lattices as described by Eq. \eqref{latticebg}. In this case, as pointed out in \cite{Andrade:2017leb}, the perturbation form in Eq. \eqref{eq_per1} is no longer applicable since now the background exhibits the periodic structure with wavenumber $k$. Instead, one needs to decompose the wave number $p$ into $\tilde{p}+nk$, where $n$ is an integer.
Specifically, we consider the linear perturbations of $B$ and $\psi$ which take the following form
\begin{equation}\label{eq:per2}
  \begin{aligned}
    \delta \psi & =\delta \psi^{\tilde{p}}(x,z) e^{i \tilde{p} x}, \\
    \delta B    & =\delta B^{\tilde{p}}(x,z) e^{i \tilde{p} x},
  \end{aligned}
\end{equation}
where $\tilde{p}$ represents the wavenumber of CDW,
ranging
from 0 to $k$. In this case, one notices that the perturbative modes with a given wavenumber $\tilde{p}$, namely $\delta\psi^{\tilde{p}}$ and $\delta B^{\tilde{p}}$, have to be $x$-dependent as well such that the resultant equations of linear perturbation become partial differential equations, which can be compactly expressed as,
\begin{equation}\label{eq:o2}
  \mathcal{O}^{\tilde{p}}(z, x)
  \left(\begin{array}{l}
      \delta \psi^{\tilde{p}} \\
      \delta B^{\tilde{p}}
    \end{array}\right)=0.
\end{equation}
The operator $\mathcal{O}^{\tilde{p}}(z, x)$
can again be expressed in matrix form after discretization. In parallel, for a given background with lattices characterized by the wave number $k$ and amplitude $\lambda$, we drop down the temperature $T$ and the instability is signaled by the existence of real eigenvalues of $\tilde{p}$ in Eq. \eqref{eq:o2}. Then we obtain a phase diagram in $ \tilde{p}-T$ plane for the background with a specified lattice. The dome below the curve represents the unstable region where CDW may emerge.

\begin{figure} [h]
  \center{
    \includegraphics[width=0.32\textwidth]{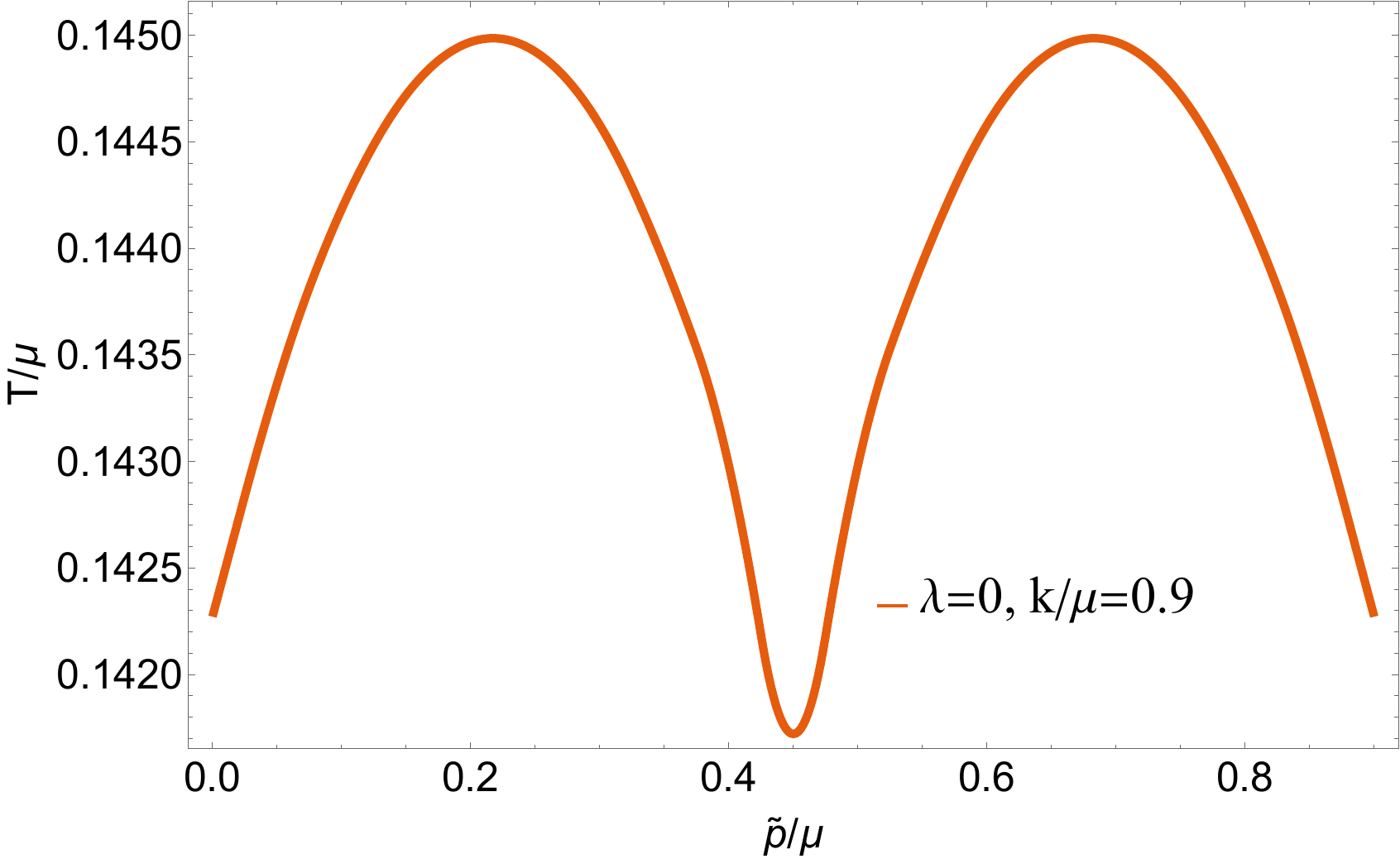}
    \includegraphics[width=0.32\textwidth]{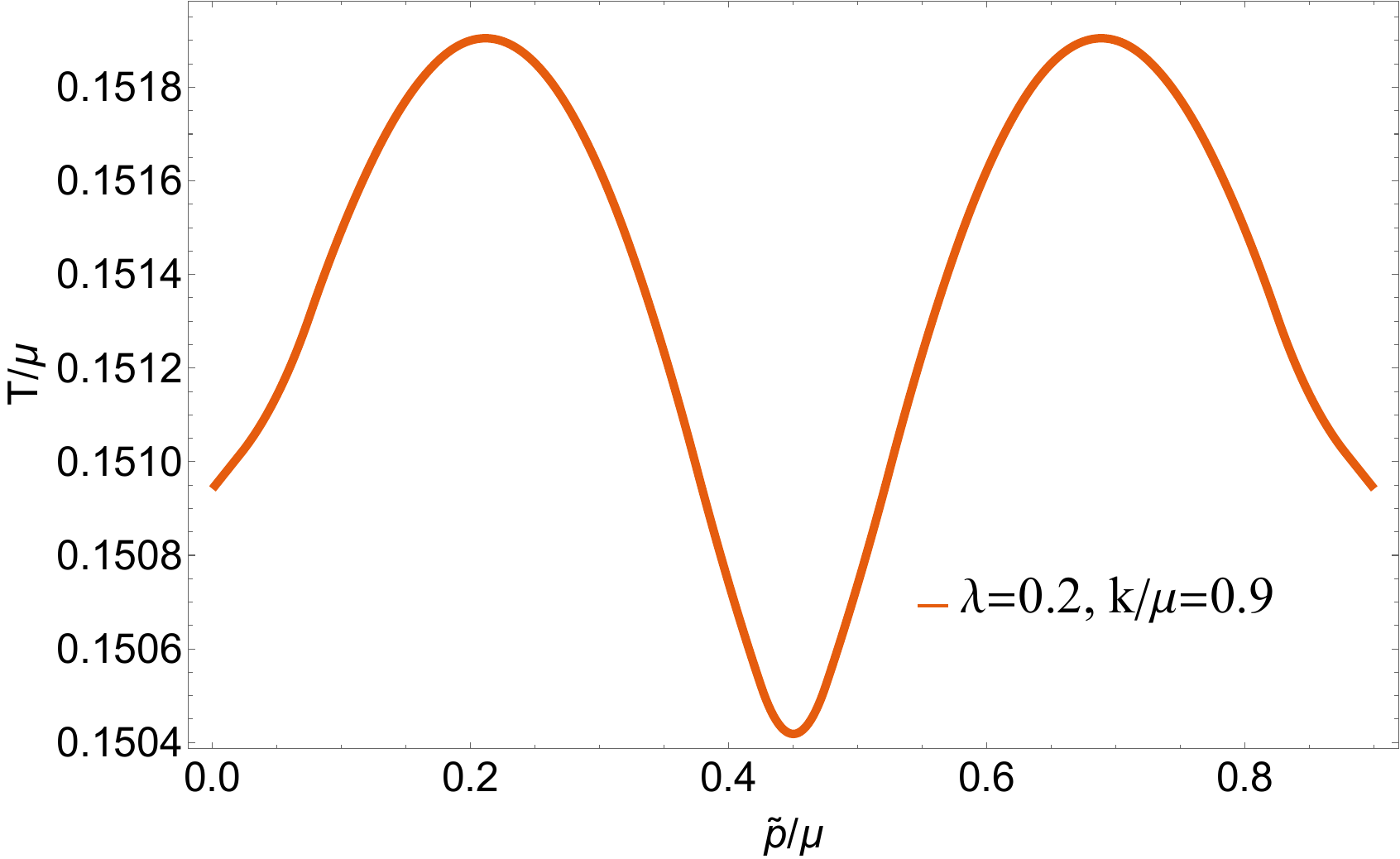}
    \includegraphics[width=0.32\textwidth]{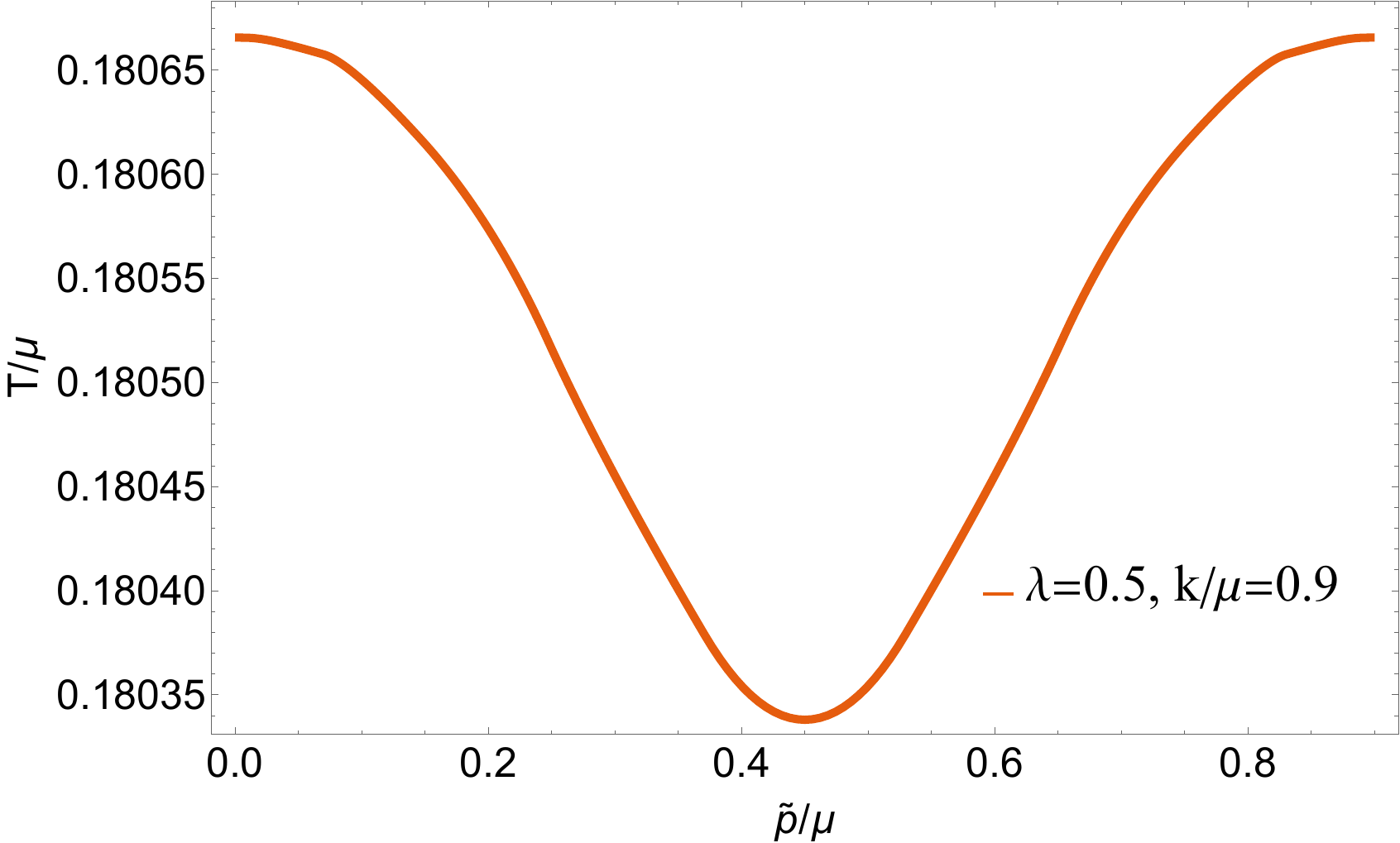}
    \includegraphics[width=0.32\textwidth]{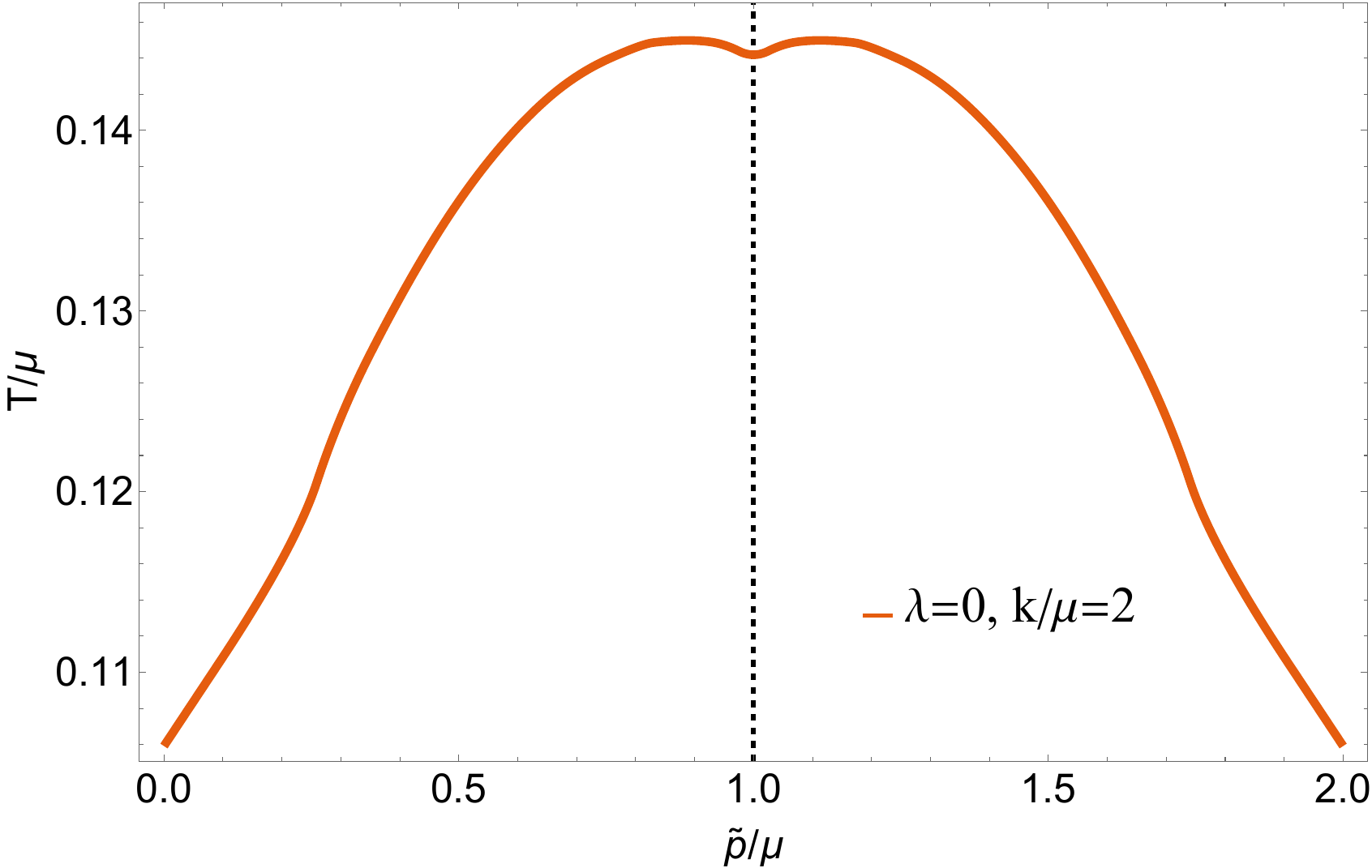}
    \includegraphics[width=0.32\textwidth]{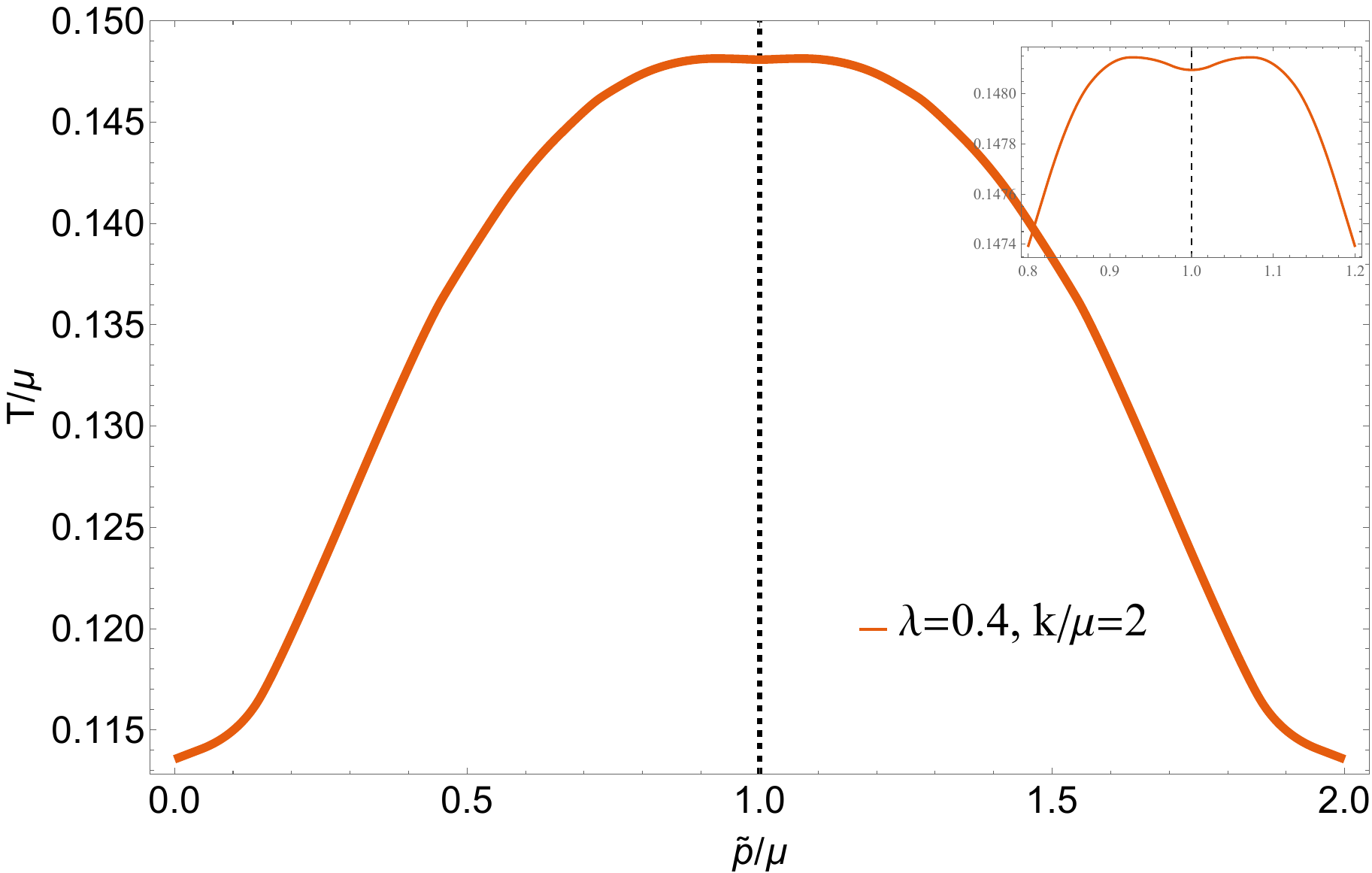}
    \includegraphics[width=0.32\textwidth]{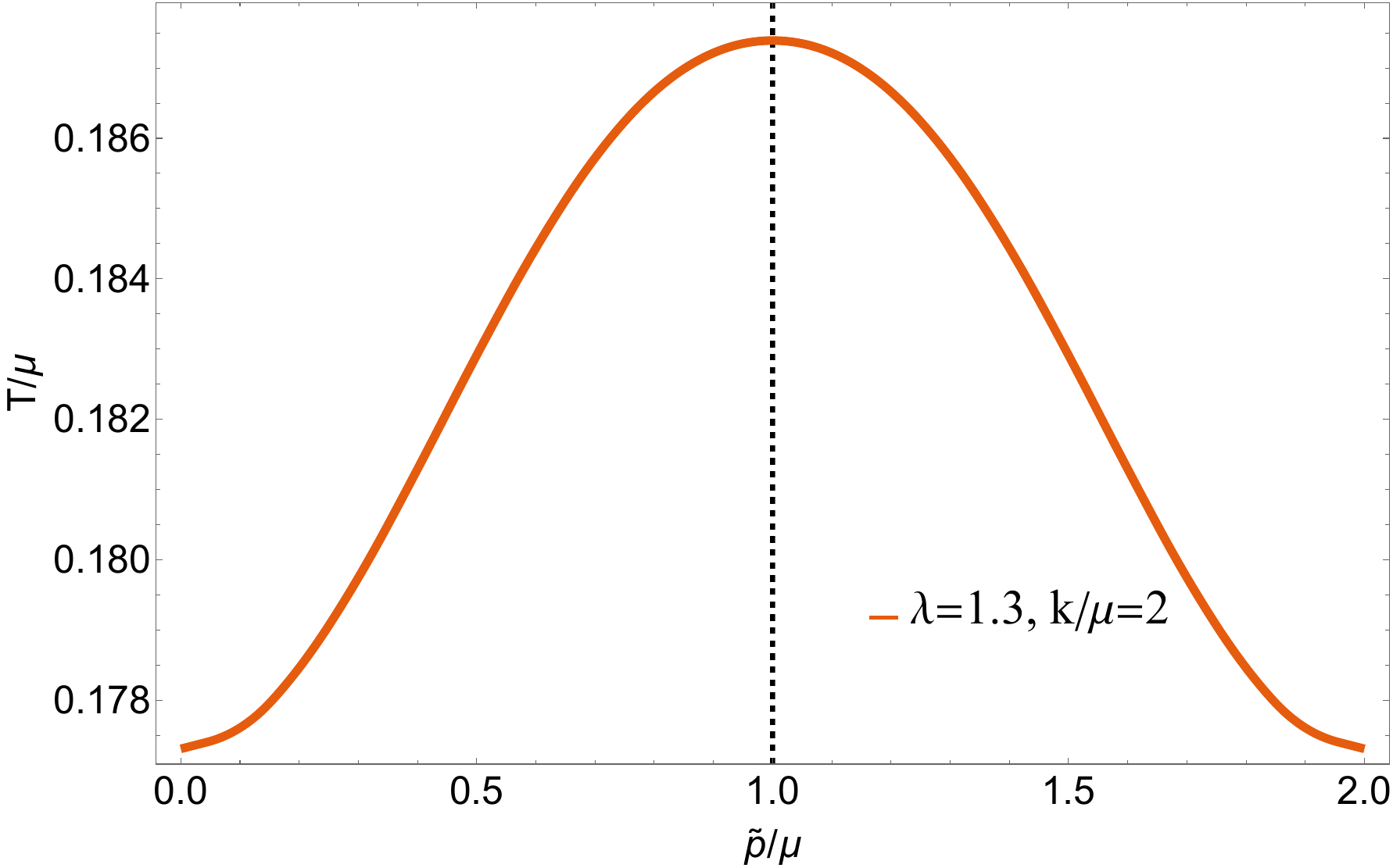}
    \caption{\label{figPT}
      The change of the phase diagram with the lattice amplitude. The first row demonstrates the lock-in process for $k/\mu=0.9$  with the commensurate value $\tilde{p}_c/k=1/1$; while the second row demonstrates the lock-in process for $k/\mu=2.0$  with the  commensurate value $\tilde{p}_c/k=1/2$.
    }}
\end{figure}

Now, we turn to the main task in this section, namely the observation of the lock-in effect in commensurate state. We fix the wavenumber $k$ of the ionic lattice, and illustrate the change of the phase diagram with the increase of the lattice amplitude. Without loss of generality, we firstly demonstrate the change of phase diagram for $k/\mu=0.9$ in the first row of Fig. \ref{figPT}, where from left to right the lattice amplitude is increased from $\lambda=0$, $\lambda=0.2$ to $\lambda=0.9$. It is interesting to notice that since $k/\mu=0.9$ is smaller than $p_c/\mu=1.1$, which is the critical wavenumber of CDW without lattices, the phase diagram is dramatically changed even for $\lambda=0$, because now the system must obey the periodicity $\tilde{p}\sim \tilde{p}+nk$. This phenomenon has also  been stressed in previous work \cite{Andrade:2017leb}. Now from the plot one can read out the critical wavenumber of CDW which is about $\tilde{p}_c / \mu=0.2$  (Due to the mirror symmetry the other one is about $\tilde{p}_c / \mu=0.7$). In comparison with $p_c/\mu\approx 1.1$ in the phase diagram without lattice, this  is reasonable since  $p_c/\mu\approx 1.1 \approx  \tilde{p}_c/\mu+k /\mu \approx 0.2+0.9$. While the critical temperature with $\lambda =0$ is almost the same as the one without lattices. Then, with the increase of the amplitude as illustrated in the middle plot with $\lambda=0.2$, we find the tips of the curve move towards the edges of the plot, and when the amplitude becomes larger, as illustrated in the right plot with $\lambda=0.9$, we find the tips arrive at the endpoints of the curve. If one increases the amplitude further, we find the tip of the curve, namely the critical wavenumber, is fixed at $\tilde{p}/\mu=0.9$, indicating that the period of CDW is locked in with the period of the background lattice, with a commensurate value $\tilde{p}_c/k=1/1$.

Next, we demonstrate another typical lock-in state with a commensurate value $\tilde{p}_c/k=1/2$. To simplify the analysis, we fix $k/\mu=2$, which is larger than $p_c/\mu=1.1$. The phase diagram in the $\tilde{p}-T$ plane for different values of the lattice amplitude $\lambda$ is shown in the second row of Fig. \ref{figPT}. For the case of $\lambda=0$, the phase diagram remains largely unchanged compared to the phase diagram without lattice. However, it exhibits a mirror symmetry with an axis at $\tilde{p}/\mu=1.0$. The tips of the curve are located around $\tilde{p}/\mu\approx 0.9$ and $\tilde{p}/\mu\approx 1.1$, resulting in $\tilde{p}/\mu=1.0$ being the bottom of a valley. As we increase the lattice amplitude, as shown in the middle plot with $\lambda=0.4$, the tips move towards the center of the plot, and the valley becomes flatter. Finally, in the right plot with $\lambda=1.3$, $\tilde{p}/\mu=1.0$ becomes the tip of the curve, representing the location of the critical wavenumber $\tilde{p}_c$. This behavior remains unchanged even with further increases in the lattice amplitude, indicating a commensurate lock-in state with $\tilde{p}_c/k=1/2$.

It is worthwhile to point out that though the critical wavenumber is locked in when the lattice strength is strong enough, the critical temperature of the phase transition is always increasing with the lattice amplitude, as illustrated in Fig. \ref{figPT}.

\begin{figure} [h]
  \center{
    \includegraphics[width=0.6\textwidth]{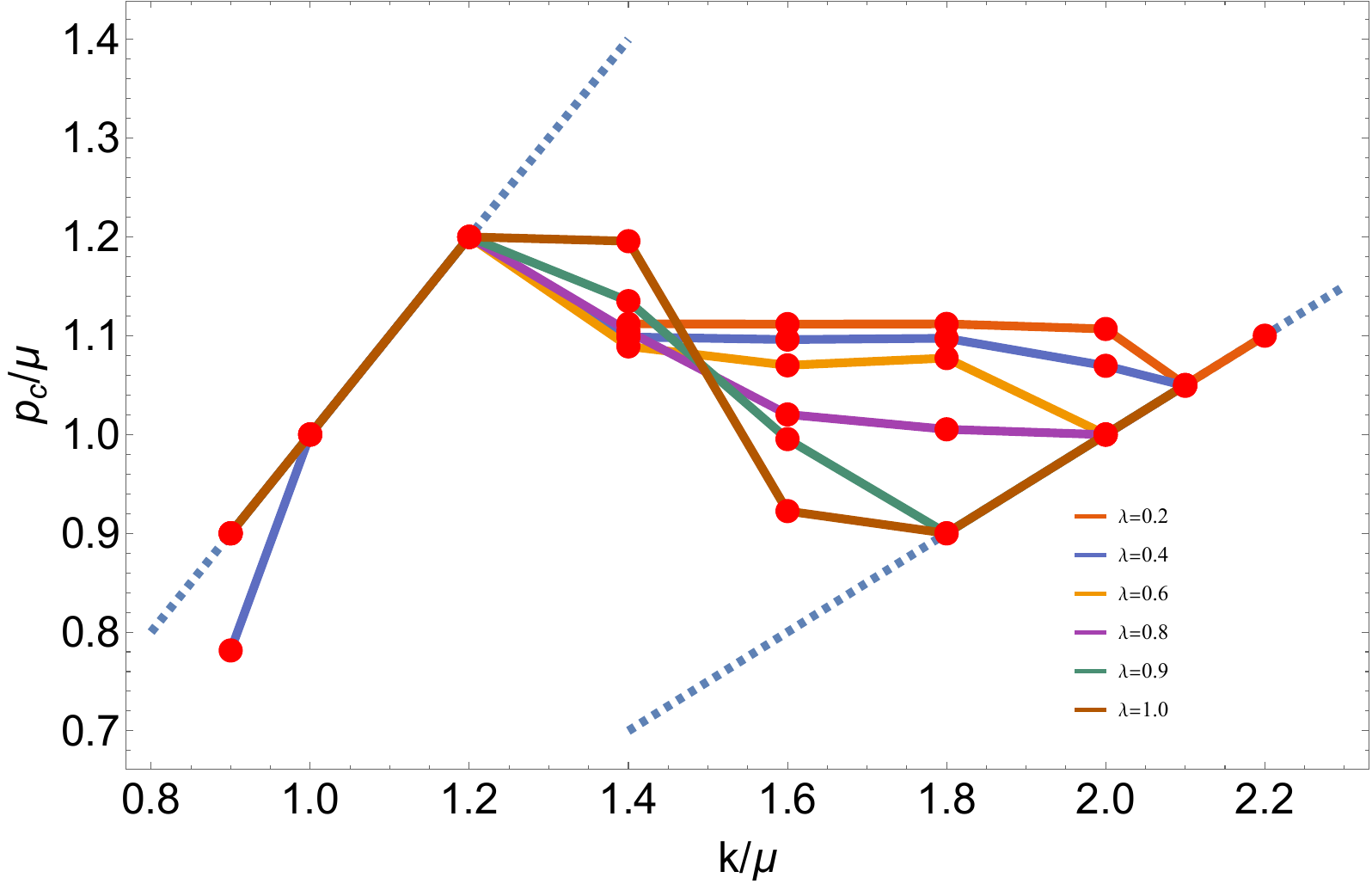}
    \caption{\label{figlock} The relationship between the critical wavenumber $\tilde{p}_c/\mu$ and $k/\mu$ for different values of lattice amplitude $\lambda$.
    }}
\end{figure}

After having presented two typical examples for commensurate lock-in state, we intend to show that lock-in effect is a general phenomenon when the lattice amplitude is large enough in this model. Thus we plot the critical curves of instability by scanning $(k,\lambda)$ space and mark the tip of each curve as the critical wavenumber, $\tilde{p}_c$. Collecting all these data, we plot Fig. \ref{figlock} to depict the relationship between the critical wavenumber $\tilde{p}_c/\mu$ and $k/\mu$ for different values of $\lambda$. Fig. \ref{figlock} shows a remarkable linear relationship between $p_c/\mu$ and $k/\mu$ at $\tilde{p}_c/k=1/1$ and $\tilde{p}_c/k=1/2$. This suggests in this model the lock-in effect is always achieved as $\lambda$ becomes large enough.

Upon concluding this section, we remark that one can obtain lock-in states with other commensurate values $\tilde{p}_c/k$, and following the analysis in \cite{Andrade:2017leb}, the devil’s ladder could also be obtained in a similar manner.

\section{The background with ionic lattices and commensurate CDW}\label{sec:numerical}

In this section, we concentrate on the construction of the background with ionic lattice and commensurate CDW by incorporating full backreaction into the system. Below the critical temperature of CDW, we adopt the following ansatz to describe the background,
\begin{equation}\label{eq:ansatz}
  \begin{aligned}
    ds^{2} & = \frac{1}{z^{2}} \left[-Q_{tt}(1-z)p(z)dt^{2} + \frac{Q_{zz}dz^{2}}{(1-z)p(z)} + Q_{xx}\left(dx+z^{2}Q_{zx}dz\right)^{2} + Q_{yy}dy^{2}\right], \\
    A      & = \mu(x)(1-z)ad t, \quad B=(1-z)bd t, \quad \psi = z\chi,
  \end{aligned}
\end{equation}
where the second gauge field $B$ and the scalar field $\psi$ are turned on. In contrast to the previous work on this model \cite{Ling:2014saa}, here we treat gauge field $A$ rather than $B$ as the electromagnetic field since we are very concerned with the distinct behavior of the conductivity before and after phase transition. Above the critical temperature of CDW, the gauge field $B$ is turned off with zero charge density, thus its conductivity is trivial as the bad metal before the phase transition. In contrast, the gauge field $A$ is turned on and in the presence of ionic lattice, the Drude law of its conductivity can be manifestly observed above the critical temperature of phase transition, exhibiting the standard metallic behavior. For this reason, we treat $A$ as the electromagnetic field and focus on its transport behavior during the phase transition of CDW. As $z \to 0$, the asymptotic behavior of
$A_t$ and $\chi$ read as,
\begin{equation}
  \begin{aligned}
    A_{t} & = \mu(x) - \rho_a(x)z + \mathcal{O}\left(z^{2}\right), \\
    \chi  & = \chi_1(x)z + \mathcal{O}\left(z^{2}\right),
  \end{aligned}
\end{equation}
where $\rho_a(x)$ is understood as the charge density of the dual system. The expansion {of $\rho_a$(x)} with Fourier modes along x-direction is,
\begin{equation}
  \rho_a(x) = \rho_{0}^{a} + \rho_{1}^{a}\cos(kx) + \cdots + \rho_{n}^{a}\cos(nkx).
\end{equation}

Now, let us drop down the temperature below the critical temperature and find numerical solutions for the background with commensurate CDW. Following the analysis in \cite{Andrade:2017ghg}, the density of CDW, denoted as $\Delta \rho_{a}(x)$, characterizes the deviation of the charge density $\rho_{a}(x)$ from the charge density over a pure ionic lattices,
denoted as $\rho_{\text{lattice}}(x)$. Then, the density of CDW is measured by
\begin{equation}
  \Delta \rho_{a}(x) = \rho_{a}(x) - \rho_{\text{lattice}}(x).
\end{equation}
In other words, $\Delta \rho_{a}(x)$ quantifies the variation of charge density at each position $x$ compared to the pure charge density over the ionic lattices at the same temperature.

For convenience, we introduce two integers, $N_k$ and $N_p$, to denote the ratio $k/\tilde{p}_c= N_k/N_p$ for commensurate states. Specifically, we fix $k/\mu = 2$ and consider the commensurate state with $k/\tilde{p}_c = N_k/N_p = 2$. The lock-in process for this case has been demonstrated in Fig. \ref{figPT}, and from Fig. \ref{figlock} we determine that the critical value of the lattice amplitude for lock-in is about $\lambda \approx 0.6$.

First of all, we numerically find that the density of CDW $\Delta \rho_{a}(x)$ and the order parameter $\chi_1(x)$ take the following form when expanded with Fourier modes
\begin{equation}
  \begin{aligned}
    \Delta \rho_{a}(x) & = \Delta \rho_{0}^{a} + \Delta\rho_{2}^{a}\cos(2\tilde{p}_cx) + \cdots \quad \text{(only even orders exist)},     \\
    \chi_1(x)          & = \chi_{1}^{1}\cos(\tilde{p}_cx) + \chi_{1}^{3}\cos(3\tilde{p}_cx) + \cdots \quad \text{(only odd orders exist)}.
  \end{aligned}
\end{equation}
We remark that this result is true for both lock-in states and unlock-in states. It reveals that only even orders of the cosine modes appear in the expression for the density of CDW $\Delta \rho_{a}(x)$, while only odd orders appear in the expression for $\chi_1(x)$. Obviously, it indicates that the CDW, namely the electronic lattice, has the same period as the ionic lattice, since $k=2\tilde{p}_c$.

\begin{figure}[h]
  \centering
  \includegraphics[width=0.45\textwidth]{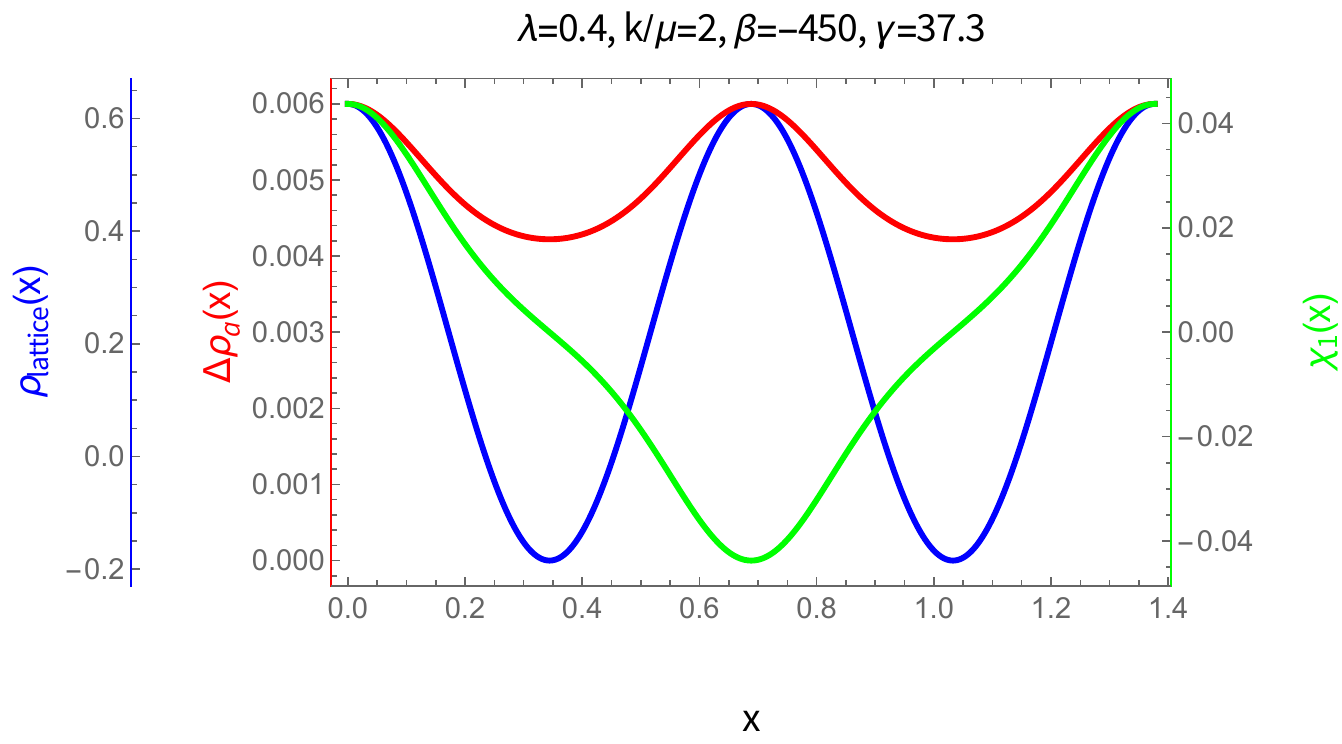}
  \includegraphics[width=0.45\textwidth]{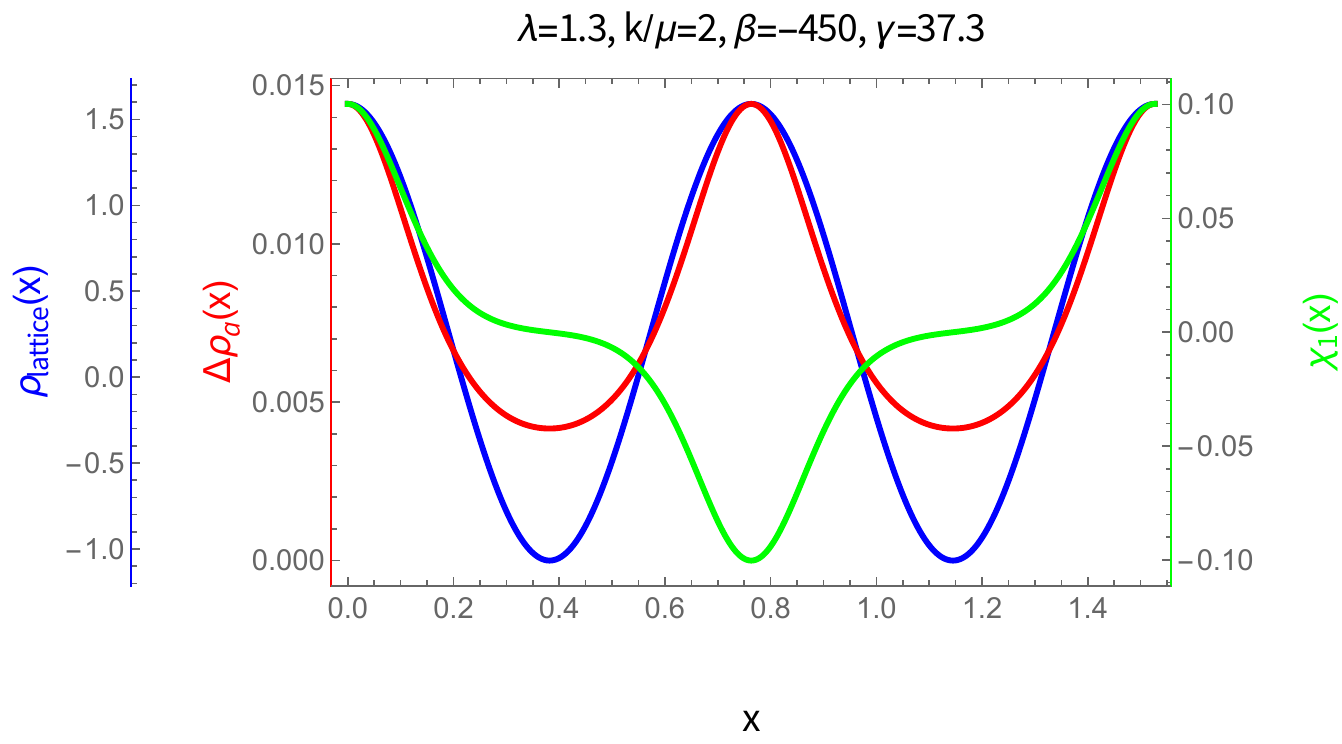}
  \caption{
    \label{fig3}
    The magnitude and periodicity of $\rho_{\text{lattice}}(x)$, $\Delta \rho_{a}(x)$, and $\chi_1(x)$ for $\lambda=0.4$ (left) and $\lambda=1.3$ (right) with $T=0.8T_c$. Notably, within one period of $\chi_1(x)$, two periods of both the lattice and the CDW can be observed.
  }
\end{figure}

In Fig. \ref{fig3}, we
plot
$\rho_{\text{lattice}}(x)$, $\Delta \rho_{a}(x)$, and $\chi_1(x)$ for two typical background solutions at the same temperature. One is the unlock-in background with $\lambda=0.4$ while the other is the lock-in background with $\lambda=1.3$. This figure indicates that CDW $\Delta \rho_{a}(x)$ and the spontaneous structure described by $\chi_1(x)$ are synchronized and exhibit a consistent pattern in the sense that $\Delta \rho_{a}(x)$ becomes non-zero only in the presence of non-zero $\chi_1(x)$, and its magnitude increases with the accumulation of $\chi_1(x)$. Moreover, comparing the magnitude of $\rho_{\text{lattice}}(x)$ and $\Delta \rho_{a}(x)$ in these two backgrounds, without surprise we find both quantities become larger with the increase of the lattice amplitude $\lambda$. Notably, we point out that the proportion of CDW among the total charge density becomes  larger in the lock-in state as well. For instance, at $\lambda=1.3$, the ratio of two magnitudes is about $\Delta \rho_{a}/\rho_{\text{lattice}}\approx 0.004$, which is in contrast to the ratio $\Delta \rho_{a}/\rho_{\text{lattice}}\approx 0.002$ for $\lambda=0.4$.

\begin{figure}[h]
  \centering
  \includegraphics[width=0.45\textwidth]{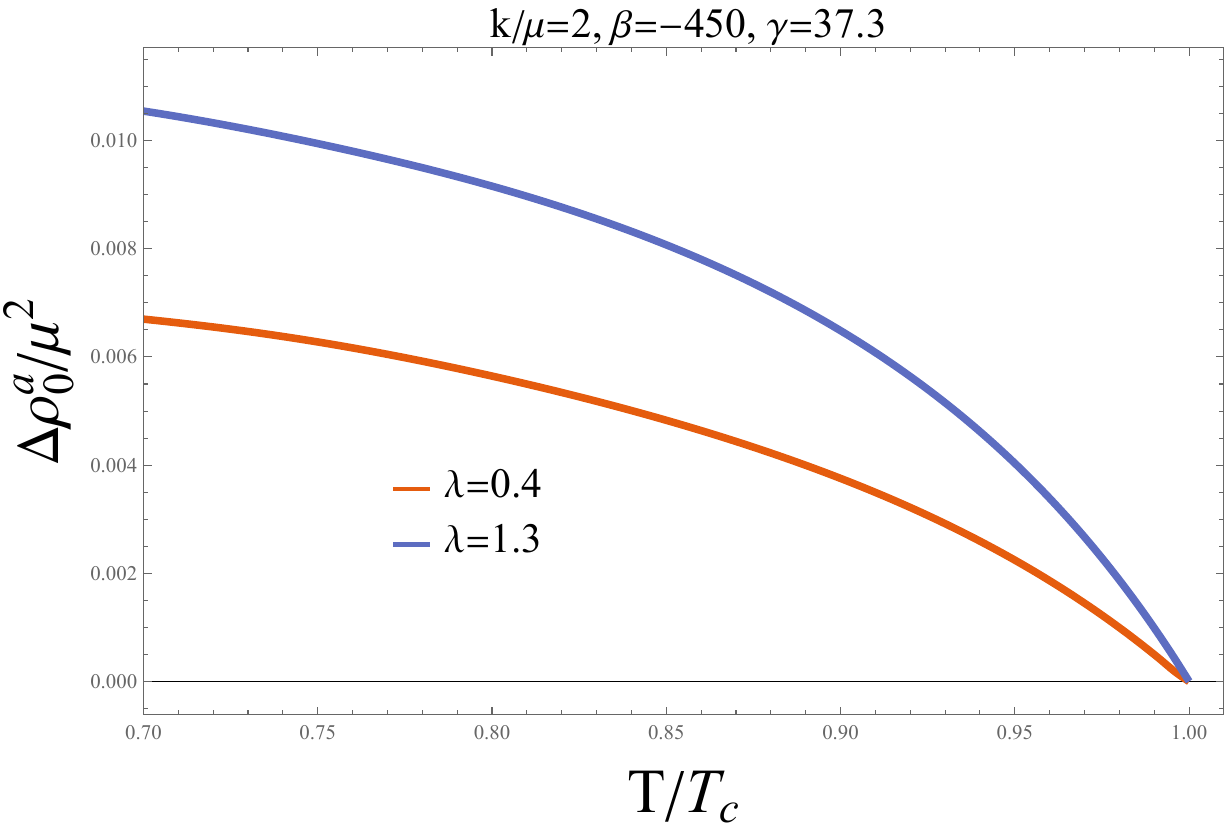}
  \includegraphics[width=0.45\textwidth]{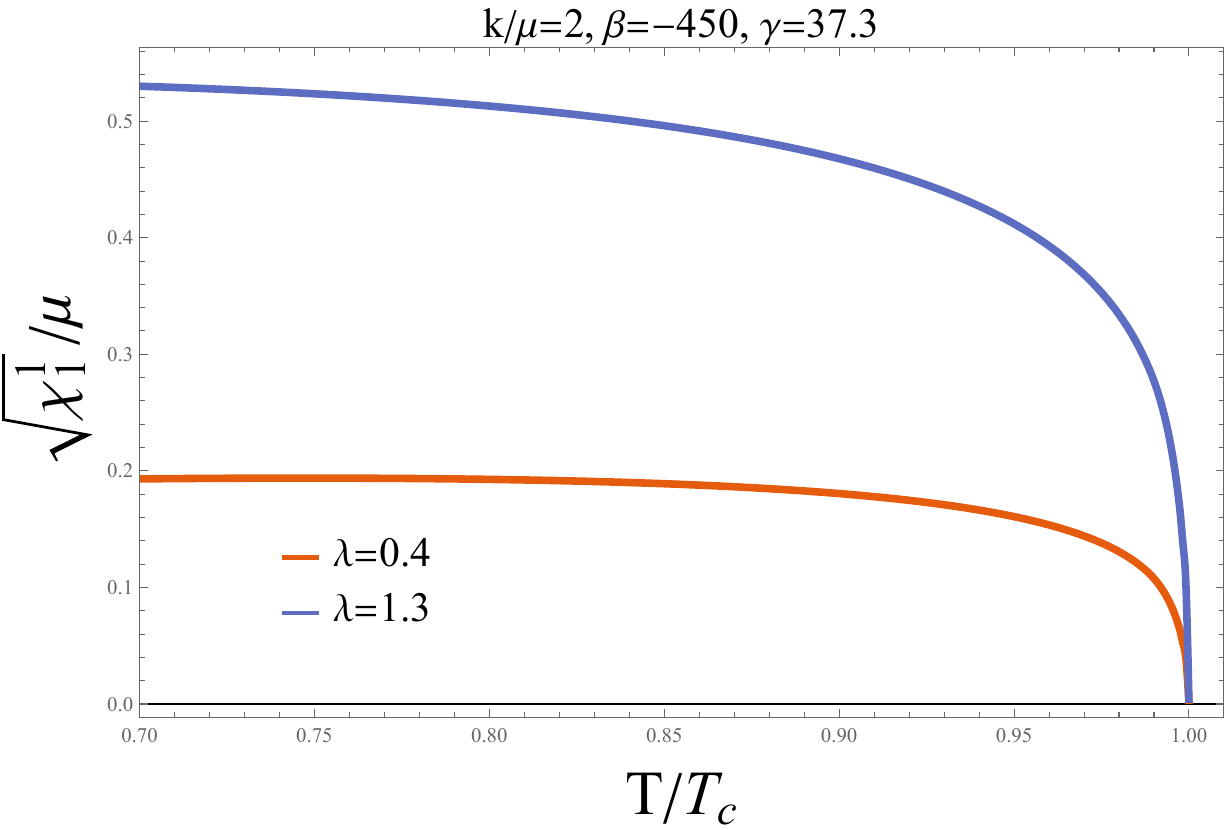}
  \caption{
    \label{fig4} The charge density of CDW  $\Delta \rho_{0}^{a}/\mu^2$ and  the order parameter $\sqrt{\chi_{1}^{1}}/\mu  $ as functions of temperature $T/T_c$.
  }
\end{figure}

Next, we
study
the variation of these quantities with temperature. In Fig. \ref{fig4}, we depict the behavior of $\Delta \rho_{0}^{a}/\mu^2$ and $\sqrt{\chi_{1}^{1}}/ \mu$ as functions of $T/T_c$. Without surprise we find both $\Delta \rho_{0}^{a}/\mu^2$ and $\sqrt{\chi_{1}^{1}}/ \mu$
increase as the temperature drops.
Notably, in a lock-in state, both quantities are much larger than those in a unlock-in state. In addition, from the behavior of $\sqrt{\chi_{1}^{1}}/ \mu$, one can justify the generation of CDW over the background with lattices is second-order phase transition.

\begin{figure}[h]
              \centering
              \includegraphics[width=0.5\textwidth]{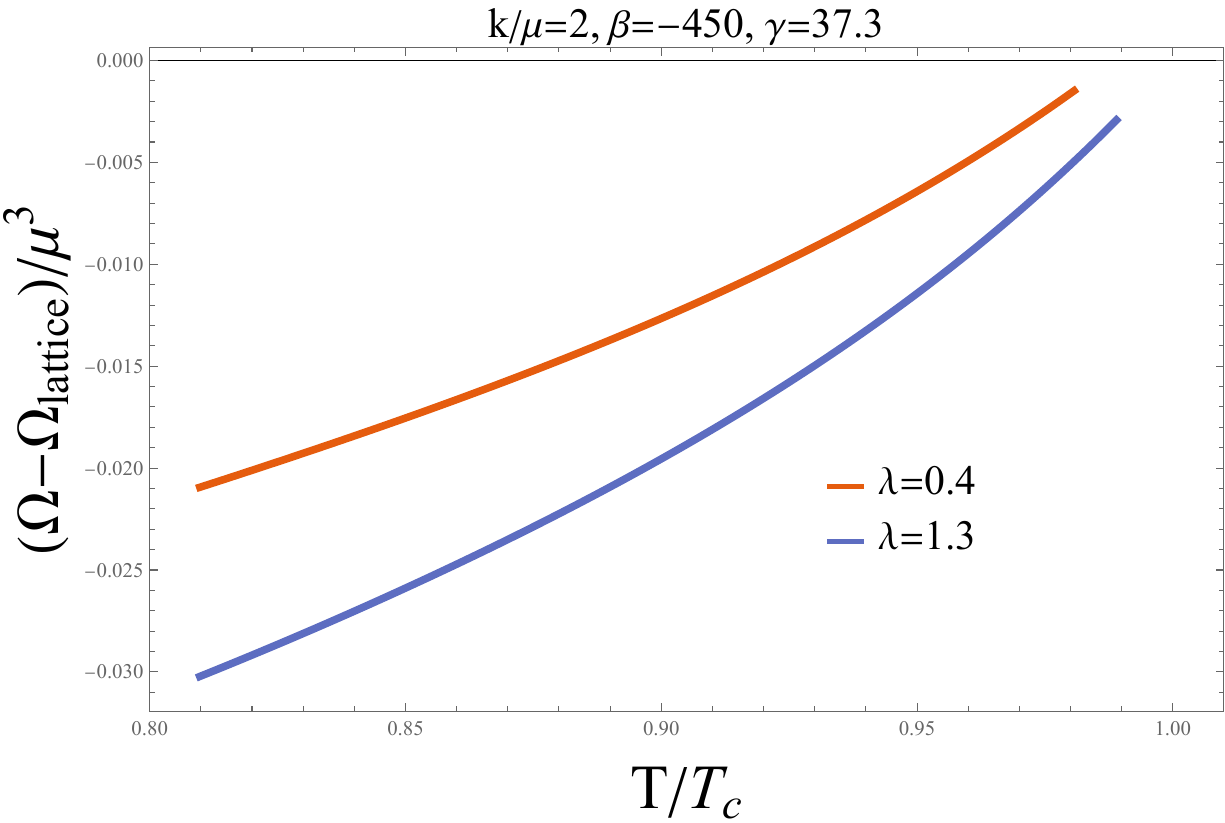} \ \hspace{0.05cm}
              \caption{
                \label{fig_free} The free energy  as functions of temperature $T/T_c$, where $\Omega$ is the free energy  of  full background for CDW, while $\Omega_{lattice}$ is the free energy  of  pure  ionic lattice.
              }
            \end{figure}

 In the end, we discuss the free energy of two different states. The averaged entropy and charge densities of our holographic model can be written as
            \begin{equation}
              \begin{aligned}
                 & S=k \int_0^{2 \pi / k} \sqrt{Q_{xx}\left(1, x\right) Q_{yy}\left(1, x\right)} d x, \\
                 & Q=\mu+\frac{k \mu}{2 \pi} \int_0^{2 \pi / k} \rho_a\left(x\right) d x.
              \end{aligned}
            \end{equation}
            Thus, the average free energy density is given by
            \begin{equation}
              \begin{aligned}
                \Omega & =M-\mu Q-T S,                                                                               \\
                M      & =2+\frac{\mu^2}{2}-\frac{3 k}{2 \pi} \int_0^{2 \pi / k} Q_{tt}^{(3,0)}\left(0,x\right) d x.
              \end{aligned}
            \end{equation}
            In Fig.\ref{fig_free}, we plot the average free energy density difference between the full background of CDW and  pure  ionic lattice background as function of the temperature. We can observe that the relative free energy of lock-in state is more stable than that of no-lock-in.

\section{The optical conductivity of CDW in commensurate state}\label{sec:optical}
To further reveal the features of lock-in states, we study the optical conductivity in the commensurate state. Specifically, we analyze the transport properties in the commensurate state where the ratio of wave numbers is $k/\tilde{p}_c = N_k/N_p = 2$.
For this purpose, we consider the following linear perturbations,
\begin{equation}
  g_{\mu \nu}=\bar{g}_{\mu \nu}+\delta g_{\mu \nu}, \quad A_\mu=\bar{A}_\mu+\delta A_\mu, \quad B_\mu=\bar{B}_\mu+\delta B_\mu, \quad \psi=\bar{\psi}+\delta \psi.
\end{equation}
We express these perturbations as deviations from the background solutions, denoted by quantities with an overbar.
We focus on the optical conductivity along the $x$ direction, as the spatial modulation is only applied in this direction. Consequently, we selectively activate $\delta A_x$ while keeping $\delta A_y$ at 0. To incorporate time-dependence, we introduce the factor $e^{-i\omega t}$ to all perturbations. At $z = 0$, we set the asymptotic behavior of $\delta A_x$ as follows,
\begin{equation}
  \delta A_x=\left(1+j_x(x) z+\cdots\right) e^{-i \omega t}.
\end{equation}
By solving the linear perturbation equations, we can deduce the optical conductivity expression as,
\begin{equation}
  \sigma(\omega)=4 \frac{j_x^{(0)}}{i \omega},
\end{equation}
where $j_x^{(0)}$ represents the first term of the Fourier expansion.

The numerical method for solving these partial differential equations has been discussed in literature \cite{Horowitz:2013jaa}. We will skip the details for this process and just mention that,
 we impose suitable gauge constraints, similar to those used in the background equations, to eliminate unnecessary degrees of freedom and ensure a correspondence between the number of fields and independent equations of motion\cite{Rangamani:2015hka,Andrade:2018gqk,Andrade:2017cnc}.
 To achieve this, it is necessary to consider additional equations to fix the gauge degrees of freedom, specifically by imposing the  Donder gauge condition for $\delta g_{\mu \nu}$ and Lorentz gauge conditions for $\delta A_\mu$ and $\delta B_\mu$,
 \begin{equation}
  \tau_{\nu} =\bar{\nabla}^\mu \hat{h}_{\mu \nu}=0, \quad \delta a=\bar{\nabla}^\mu \delta A_\mu=0, \quad \delta b=\bar{\nabla}^\mu \delta B_\mu=0,
\end{equation}
where $\hat{h}_{\mu \nu}=\delta g_{\mu \nu}-\delta g \bar{g}_{\mu \nu} / 2$.
The procedure to impose these gauge conditions is analogous to the method used in the DeTurck approach. For instance, to impose the de Donder gauge, we modify the Einstein perturbation equation by adding a new term:
\begin{equation}
\delta G^{H}_{\mu \nu}=\delta G_{\mu \nu}-\bar{\nabla}_{(\mu } \tau_{\nu)}
\end{equation}
This modification reveals that the Einstein perturbation equation is hyperbolic.
For the perturbation equations of gauge fields $\delta A_\mu$ and $\delta B_\mu$, we incorporate a gauge-fixing term comprising the combination of $\mathcal{F}(x,z)\bar{\nabla}_{\mu} \delta a$ and $\mathcal{G}(x,z)\bar{\nabla}_{\mu} \delta b$, where $\mathcal{F}(x,z)$ and $\mathcal{G}(x,z)$ are combination functions that are relevant to the background field. This inclusion ensures that the gauge field equations are in a hyperbolic form.

Next, we present our numerical results for the conductivity in unlock-in case and lock-in case separately. In each case, we intend to observe the behavior of conductivity in two distinct ranges of temperature. The first range is close to the critical temperature where the metal-insulator phase transition occurs. We will reveal the temperature behavior of DC conductivity during the course of phase transition. The second range is well below the critical temperature, where we will reveal the emergent process of the pseudogap and finally demonstrate its saturation at lower temperatures. To ensure the robustness of the conductivity computations, we also examined the quasi-normal modes of the system by analyzing the behavior of $\sigma(\omega)$ across the complex frequency range, as detailed in Appendix \ref{app:qnm-analysis}.

\subsection{Optical Conductivity in the Commensurate Unlock-in State}
\begin{figure} [h]
  \center{
    \includegraphics[width = 0.45 \textwidth]{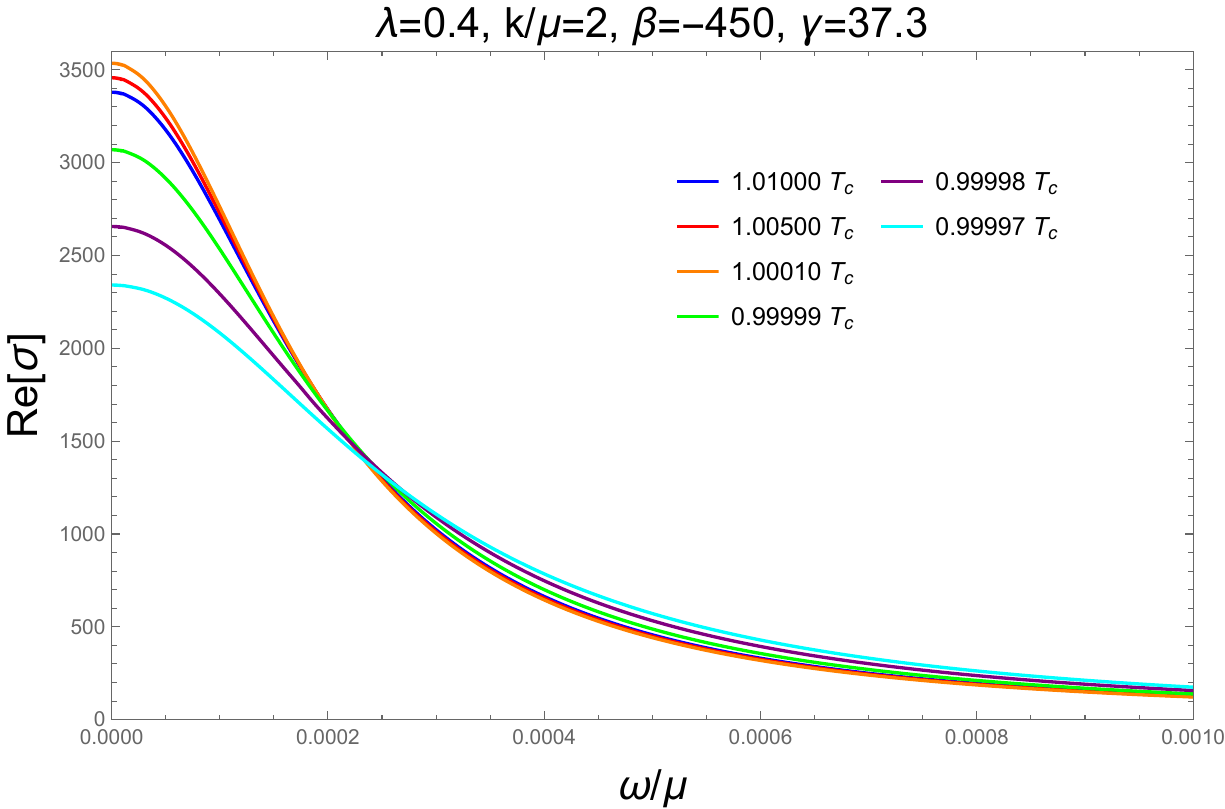}
    \includegraphics[width = 0.45 \textwidth]{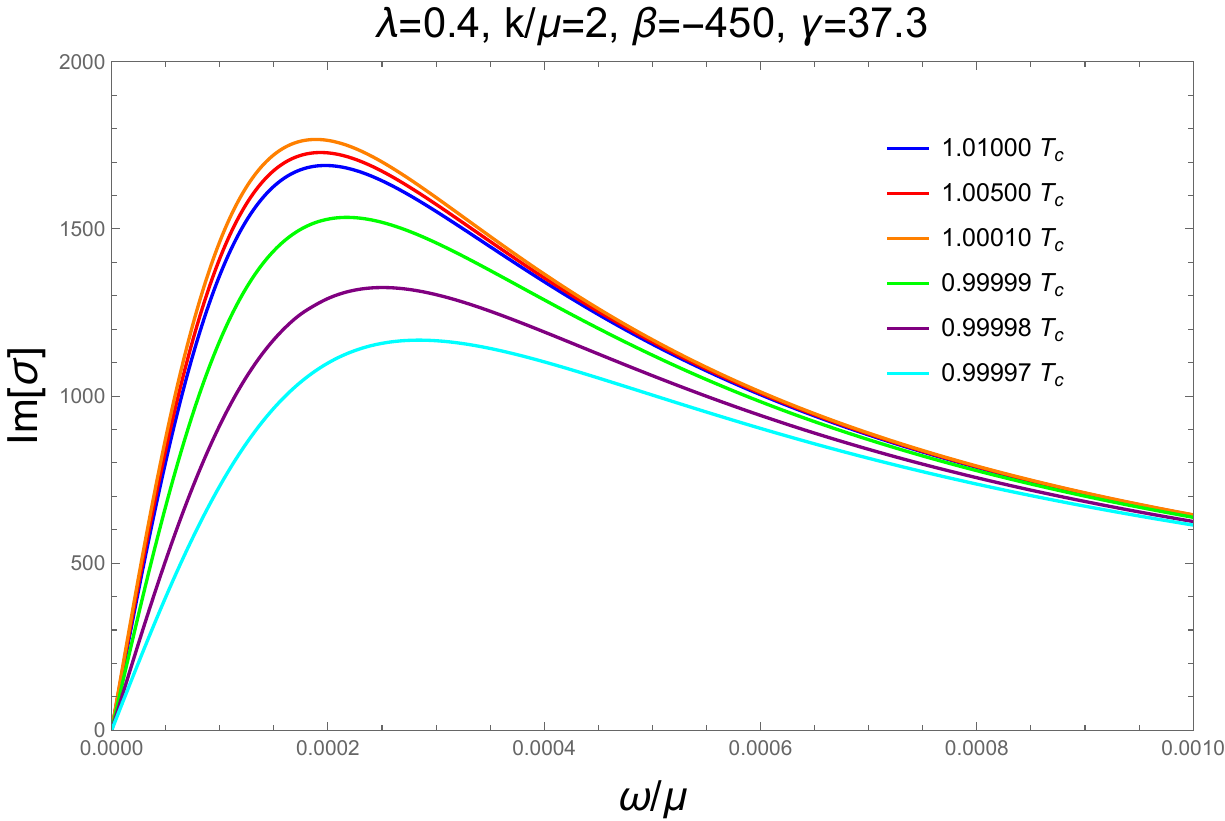}
    \caption{\label{figsigmacomm04tc} The optical conductivity as a function of the frequency at temperature close to $T_c$ for a commensurate unlock-in state with $\lambda=0.4$.}}
\end{figure}

\begin{figure} [h]
  \center{
    \includegraphics[width = 0.45 \textwidth]{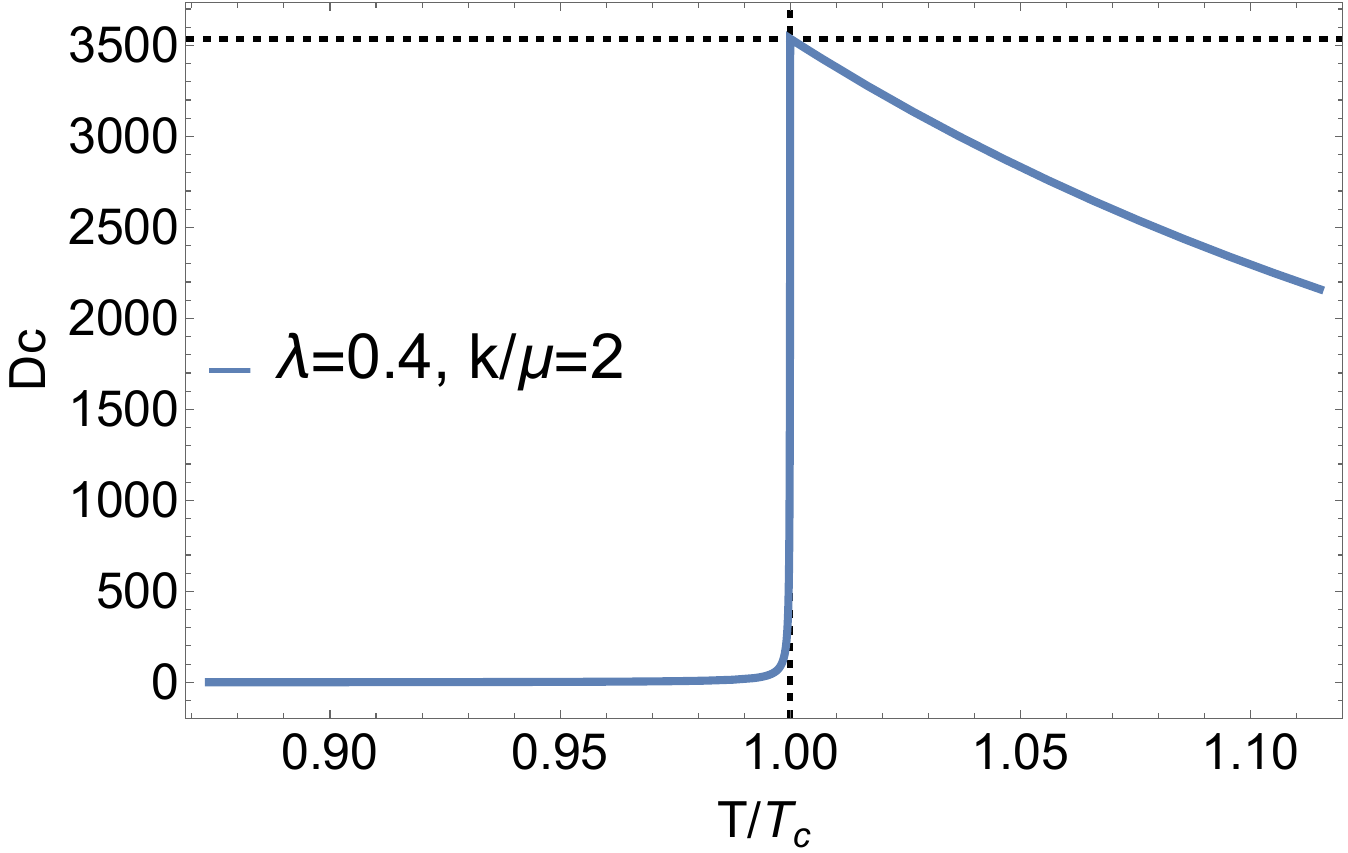}
    \caption{\label{figsigmacomm04Dc} The Dc conductivity as a function of the frequency for a commensurate unlock-in state with $\lambda=0.4$.}}
\end{figure}

In this subsection, we analyze the optical conductivity in the unlock-in case with $\lambda = 0.4$ and $k/\mu = 2$. We first examine the behavior of the optical conductivity near the critical temperature ($T_c$). As shown in Fig. \ref{figsigmacomm04tc}, above $T_c$, the system displays metallic behavior, with an increase in DC conductivity as temperature decreases. The AC conductivity can be fitted well using the standard Drude formula. Below $T_c$, the system undergoes a transition into an insulating state, with a rapid decrease in DC conductivity as temperature decreases (see Fig. \ref{figsigmacomm04Dc}). Therefore, a clear metal-insulator transition occurs at the critical temperature $T_c$. It is worth mentioning that we compute the DC conductivity by employing the value of AC conductivity at zero frequency for simplicity.

\begin{figure} [h]
  \center{
    \includegraphics[width = 0.45 \textwidth]{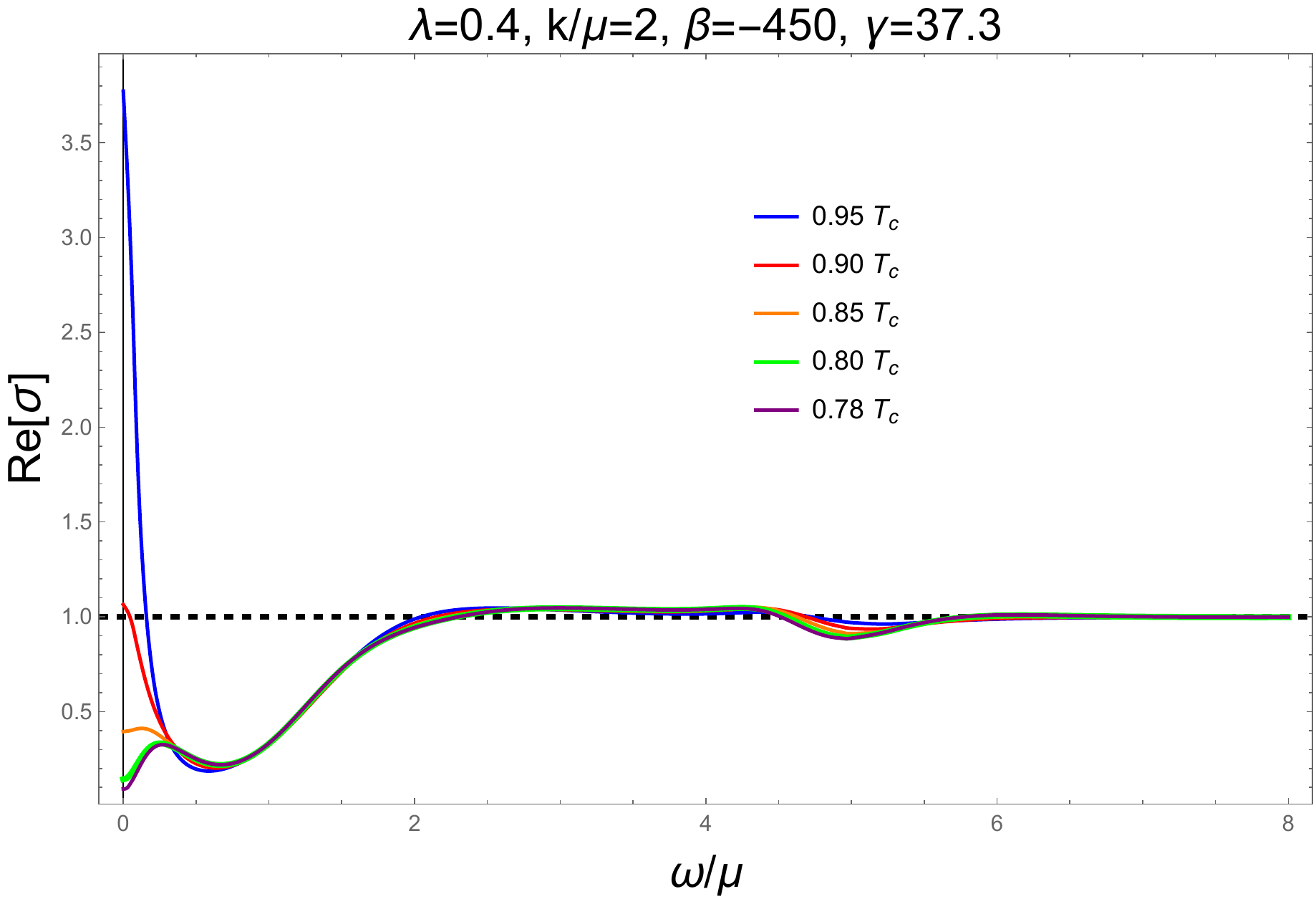}
    \includegraphics[width = 0.45 \textwidth]{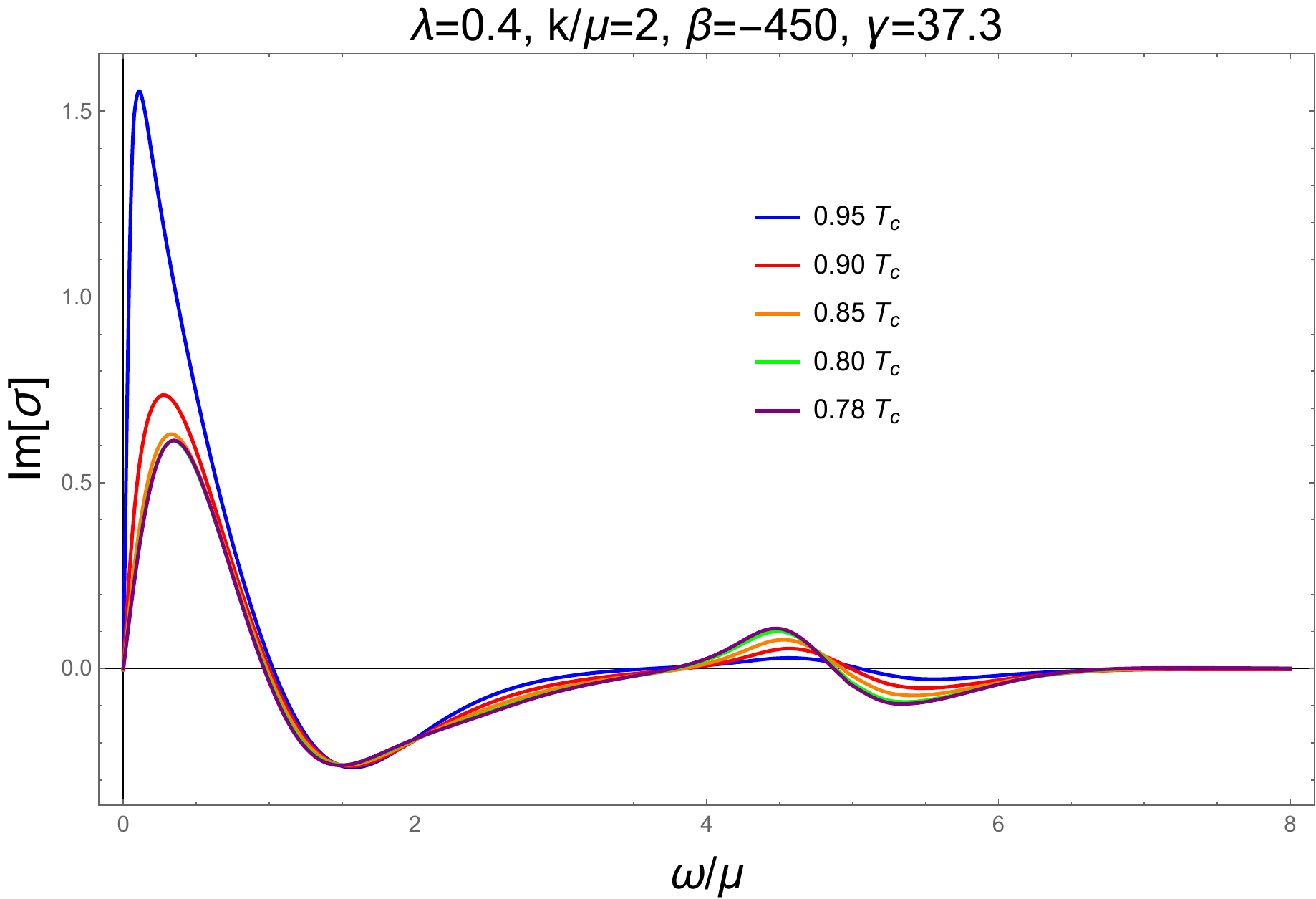}
    \caption{\label{figsigmacomm04}  The optical conductivity further away from the critical temperature for $\lambda=0.4$. }}
\end{figure}

\begin{figure} [h]
  \center{
    \includegraphics[width = 0.45 \textwidth]{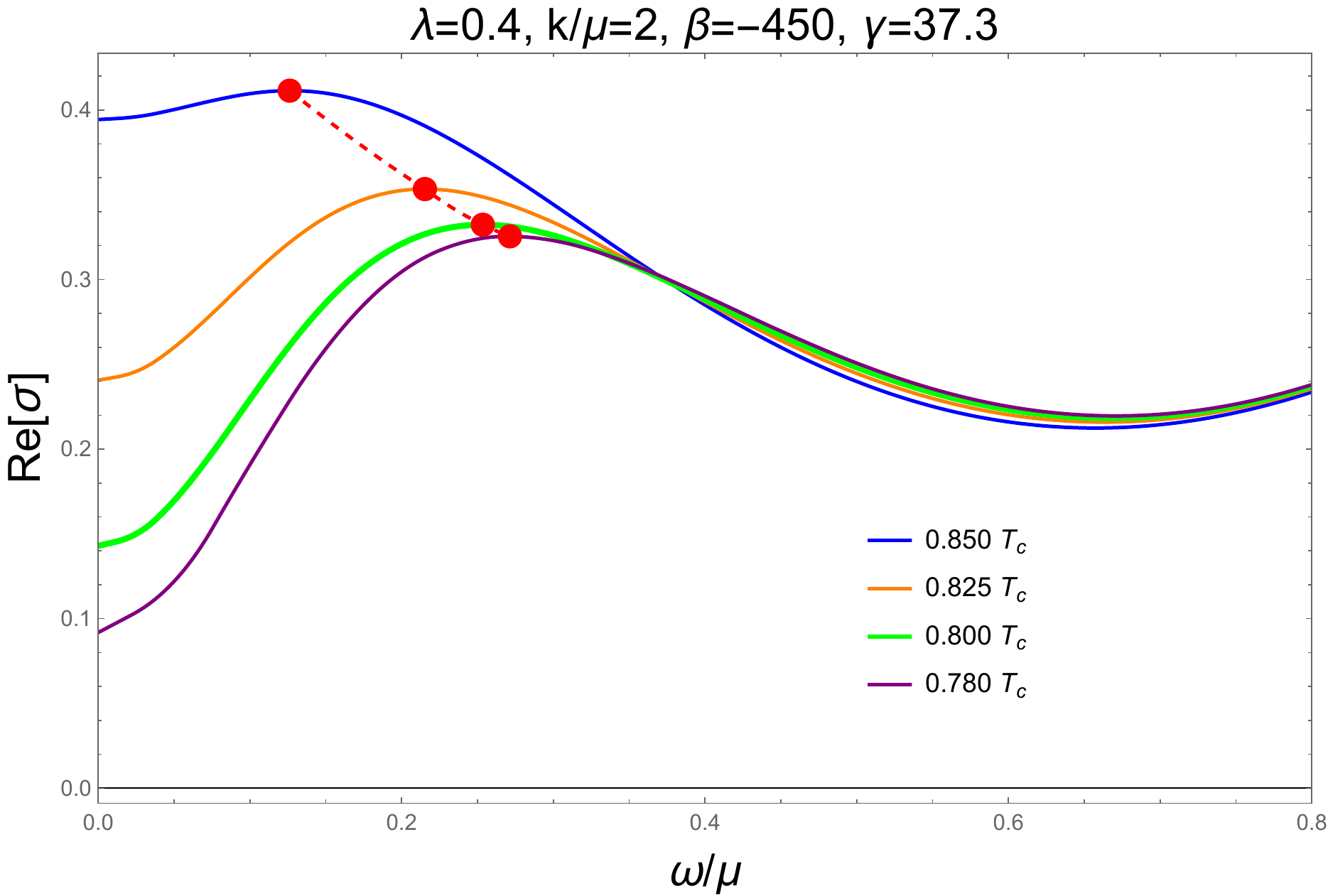}
    \caption{\label{figsigmacomm04v2}  The optical conductivity further away from the critical temperature for $\lambda=0.4$ in low range of frequency. }}
\end{figure}

\begin{figure}
  \centering
  \includegraphics[width=0.5\textwidth]{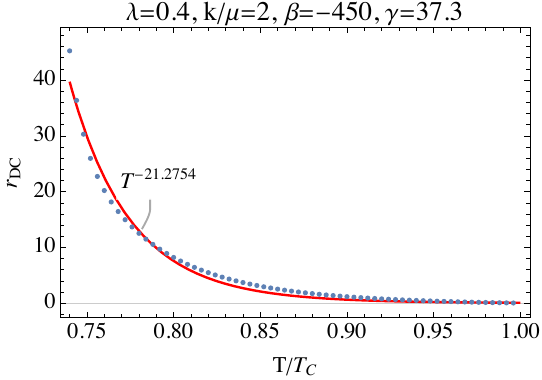}
  \caption{DC resistivity ($r_{\text{DC}}$) as a function of the reduced temperature ($T/T_c$). Notably, an approximate power-law behavior of $T^{-21.2754}$ is observed, indicating an algebraic increase of $r_{\text{DC}}$ with decreasing temperature.}
  \label{fig:rdc04}
\end{figure}

Next, we demonstrate the optical conductivity far below the critical temperature.
Fig. \ref{figsigmacomm04} shows the optical conductivity at various temperatures below $T_c$. At $T \approx 0.85T_c$, we observe a shift of the Drude peak from zero frequency to the pinning frequency as the temperature decreases.

Fig. \ref{figsigmacomm04v2} reveals the electrical conductivity behavior in the low-frequency range, where the Drude peak is observed to shift towards higher frequencies as the temperature decreases. Moreover, it is evident that the frequency associated with the Drude peak decreases less with further temperature reduction. This observation implies that the CDW is pinned to the background lattice, restricting its mobility. During the transition from a metallic to an insulating state, the CDW becomes fixed in specific positions on the ionic lattice, leading to the formation of a stable structure.

In the end of this subsection,
we plot the DC resistivity as a function of temperature below $T_c$ in Fig. \ref{fig:rdc04}. We observe that the DC resistivity presents a slow, almost algebraic increase, rather than an exponential one as the temperature decreases. This behavior aligns with previous findings reported in \cite{Andrade:2017ghg}, where a similar slow algebraic increase was observed in commensurate lock-in states, and is consistent with certain experimental results \cite{Boebinger:1996,Laliberte:2016,Shi:2021}. Nevertheless, it is noticed that this slow increase in Fig. \ref{fig:rdc04} does not conform precisely to a pure power function of the temperature. This observation suggests that in a commensurate unlock-in state DC resistivity exhibits the slow increase with decreasing temperature, but does not strictly follow an algebraic pattern.

\subsection{The optical conductivity in commensurate lock-in state}

In this subsection, we compute the optical conductivity in a commensurate lock-in state and compare its features with those in a commensurate unlock-in state. Specifically, we consider the transport behavior in the presence of CDW  with wavenumber $\tilde{p}_c/\mu=1.0$ over the background with ionic lattice $\lambda = 1.3, k / \mu = 2.0$.

\begin{figure} [h]
  \center{
    \includegraphics[width=0.45\textwidth]{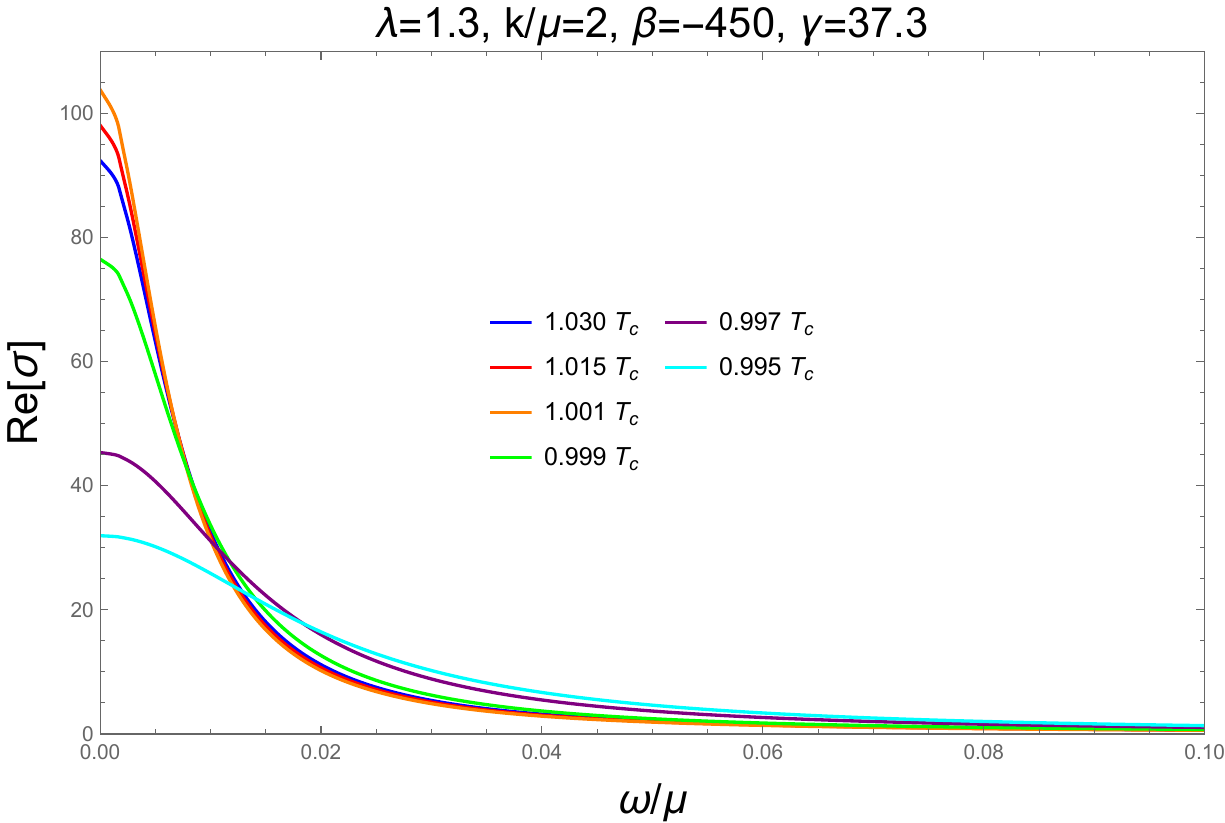}
    \includegraphics[width=0.45\textwidth]{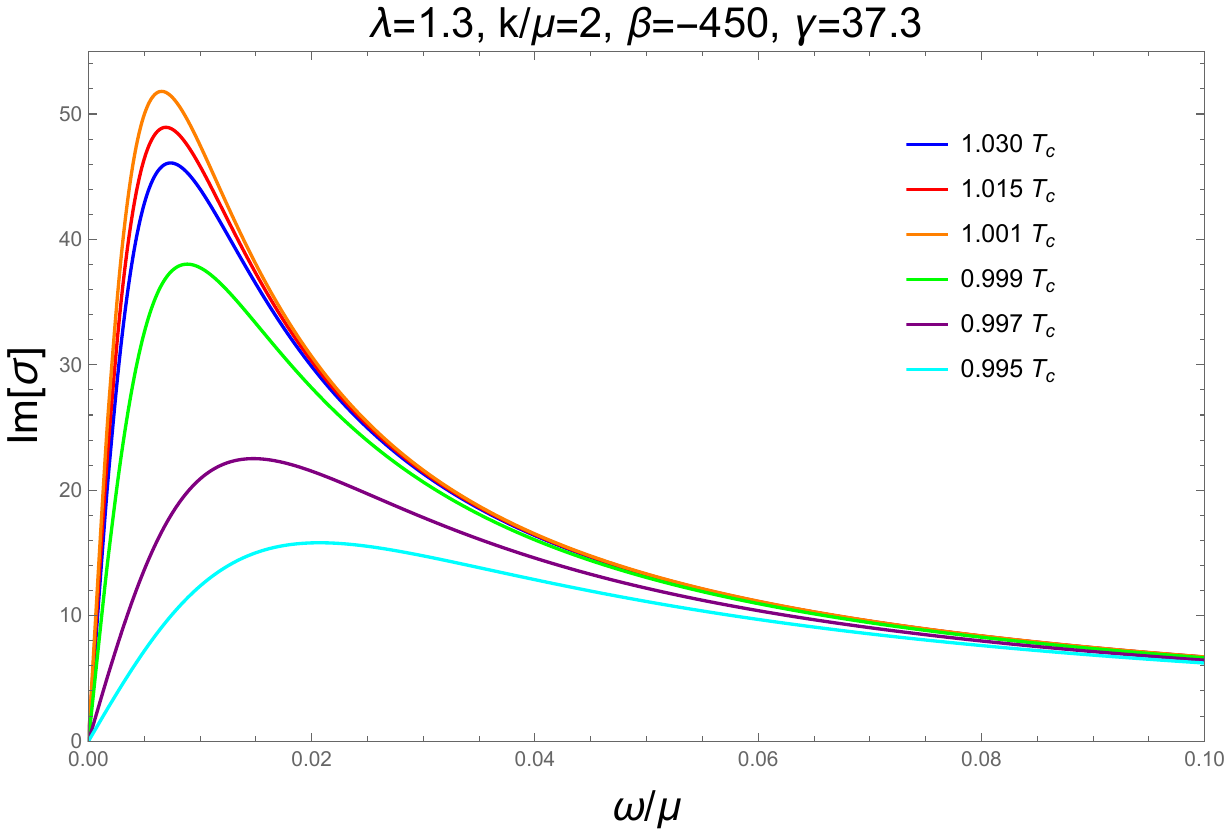}
    \caption{\label{figsigmacomm13tc} The optical conductivity at temperature close to $T_c$ for $\lambda=1.3$. }}
\end{figure}

\begin{figure} [h]
  \center{
    \includegraphics[width = 0.45 \textwidth]{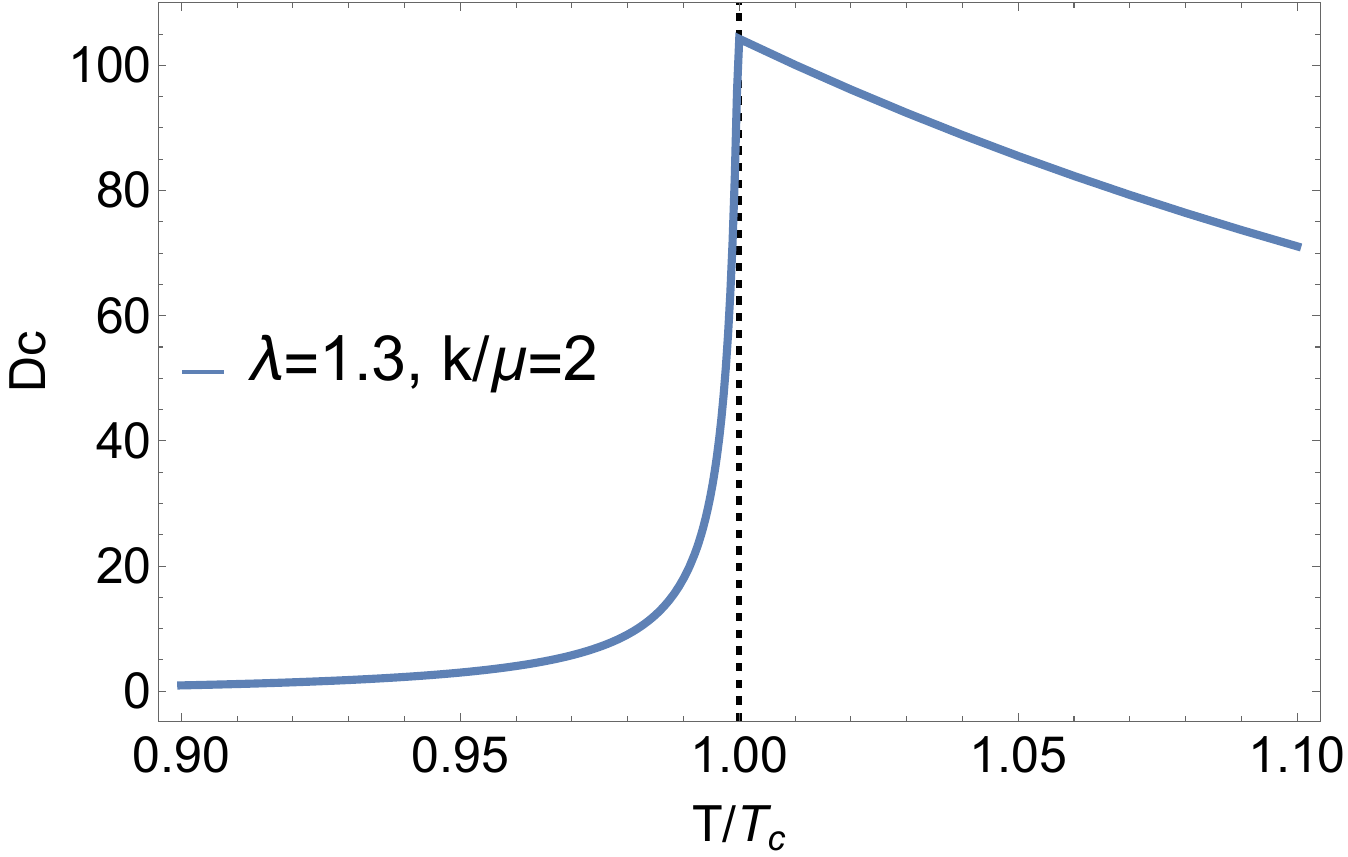}
    \caption{\label{figsigmacomm13Dc} The Dc conductivity as a function of the frequency for a commensurate unlock-in state with $\lambda=1.3$.}}
\end{figure}

Firstly, we investigate the optical conductivity at a temperature close to $T_c$, as illustrated in Fig. \ref{figsigmacomm13tc}. Similar to the unlock-in case,
we observe a metal-insulator transition
near the critical temperature since the DC conductivity exhibit distinct behavior for temperatures above $T_c$  and below $T_c$(see Fig. \ref{figsigmacomm13Dc}).
Moreover, in the lock-in state, the value of the DC conductivity is much smaller compared to the unlock-in state. This indicates that the influence of the lattice becomes stronger as the lattice amplitude increases.

\begin{figure} [h]
  \center{
    \includegraphics[width=0.45\textwidth]{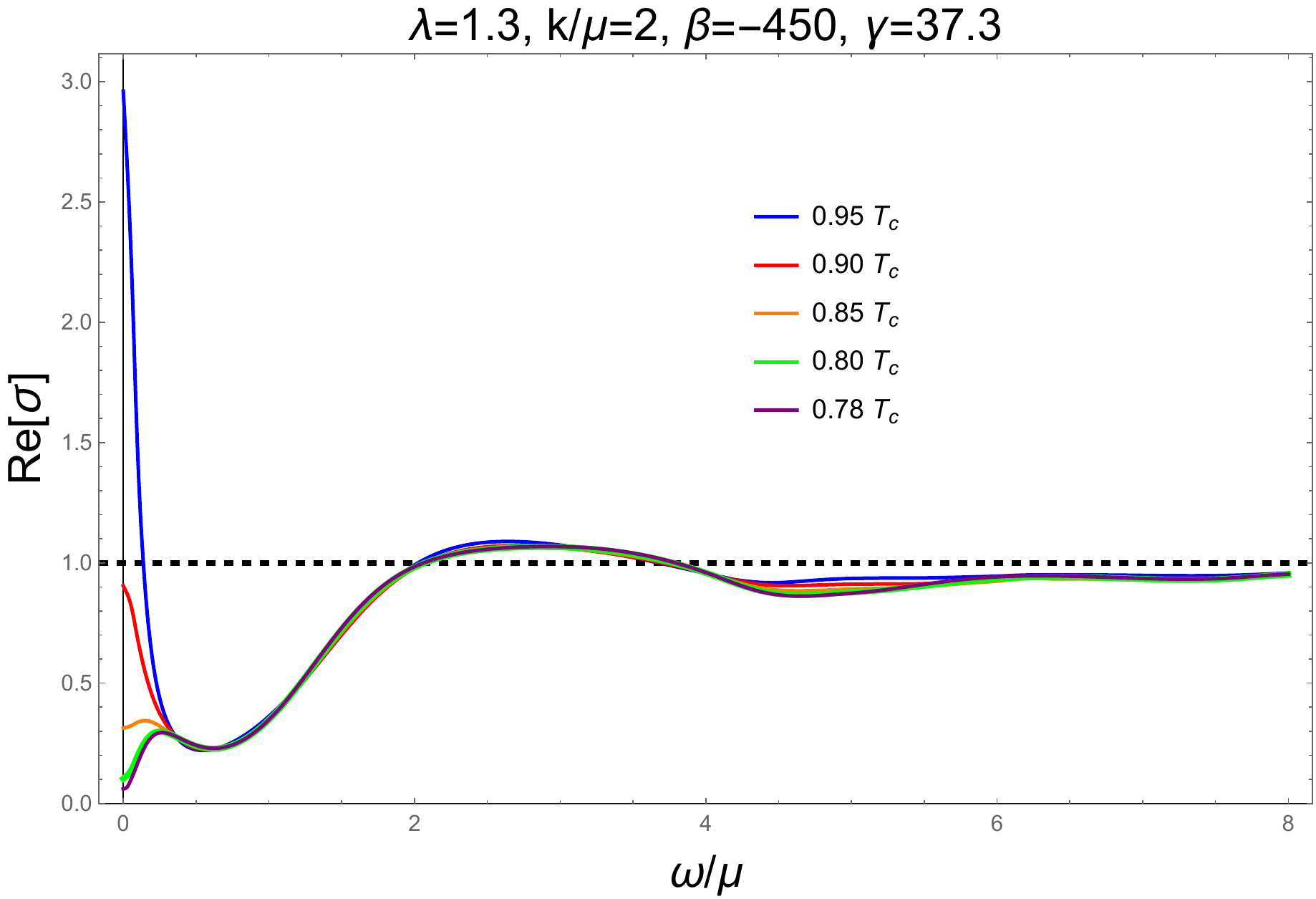}
    \includegraphics[width=0.45\textwidth]{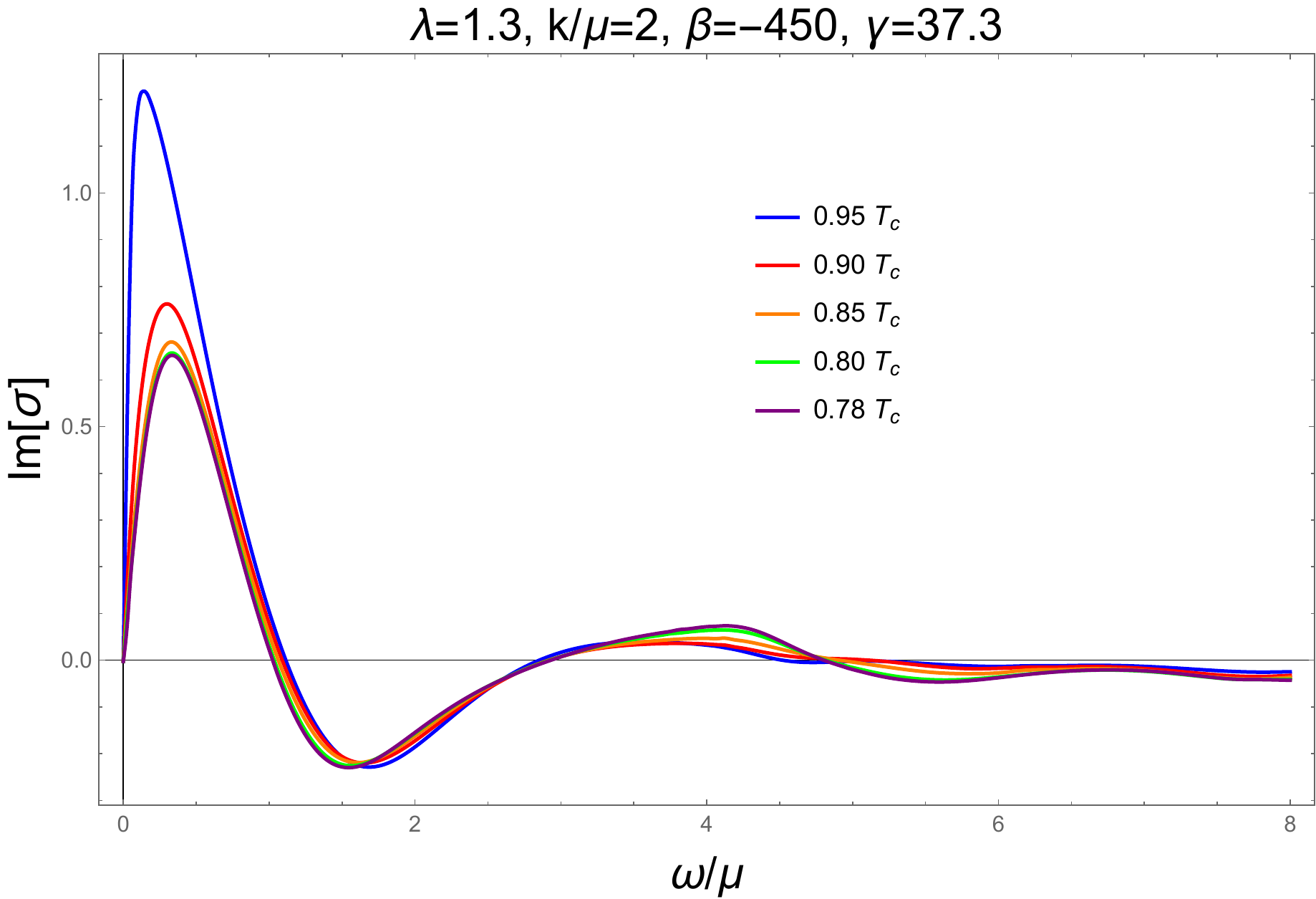}
    \caption{\label{figsigmacomm13} The optical conductivity further away from the critical temperature for $\lambda=1.3$. }}
\end{figure}

\begin{figure} [h]
  \center{
    \includegraphics[width = 0.45 \textwidth]{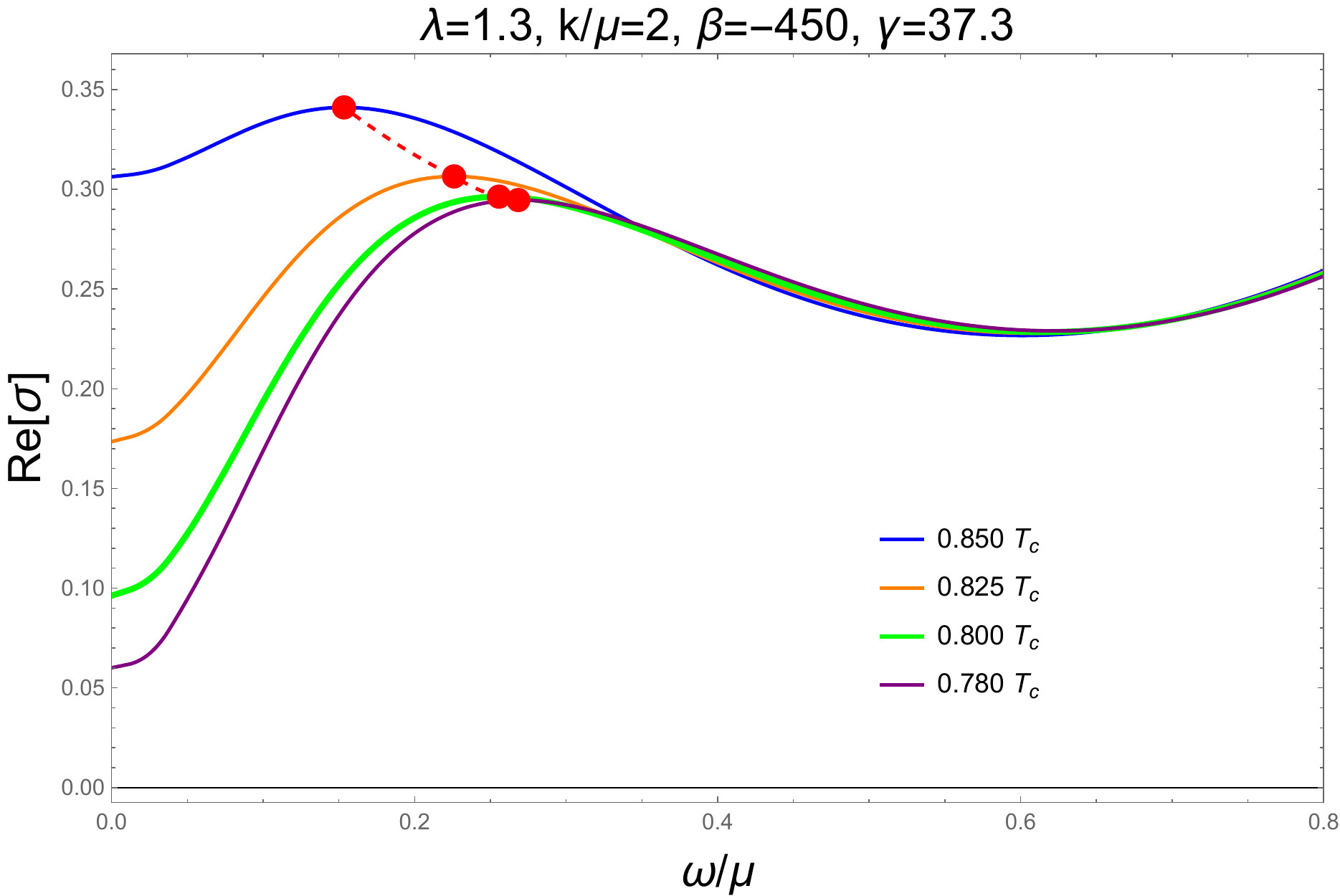}
    \caption{\label{figsigmacomm13v2}  The optical conductivity further away from the critical temperature for $\lambda=1.3$ in low range of frequency. }}
\end{figure}

Next, we proceed to compute the optical conductivity at temperatures well below the critical temperature, which is illustrated in Fig. \ref{figsigmacomm13}. Similarly,  a decline in the DC conductivity is observed as the temperature decreases, accompanied by a shift of the Drude peak from zero frequency to the pinning frequency with decreasing temperature.
 As shown in Fig. \ref{figsigmacomm13v2}, the locked state results in a stronger pinning effect, which in turn diminishes the Drude peak. This observation underscores the substantial impact of the CDW on the  electrical properties, particularly in terms of limiting electron mobility and enhancing the insulating characteristics.

\begin{figure}[h]
  \centering
  \includegraphics[width=0.5\textwidth]{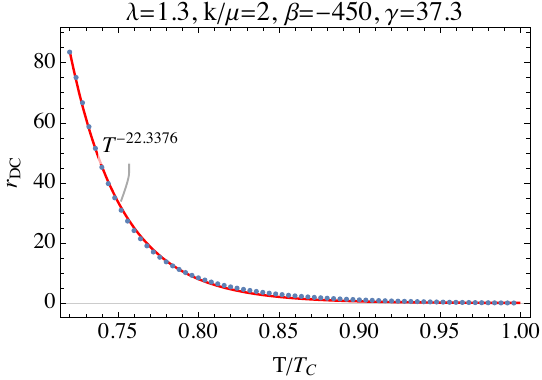}
  \caption{The DC resistivity ($r_{\text{DC}}$) as a function of the reduced temperature ($T/T_c$). Notably, a good power-law behavior of $T^{-22.3376}$ is observed, indicating an algebraic increase of $r_{\text{DC}}$ with decreasing temperature.}
  \label{fig:rdc13}
\end{figure}

 Additionally, we find that the DC resistivity in the lock-in case not only shows a slow increase with decreasing temperature, similar to the unlock-in case, but also can be accurately fitted with a power function of temperature, as shown in Fig. \ref{fig:rdc13}. This finding is in line with the results reported in \cite{Andrade:2017cnc}. The above observation suggests that in our model, both the unlock-in and lock-in cases exhibit a slow increase in DC resistivity, but only the lock-in case demonstrates the characteristic algebraic increase. This implies that the lock-in effects play a crucial role in determining the algebraic behavior of the DC resistivity.

\section{Discussion}\label{sec:dis}

In this paper, we have considered a simple holographic model to investigate the commensurate lock-in effect and compared the transport properties in lock-in state and unlock-in state. This model contains two $U(1)$ gauge fields and the instability of the background is induced by the coupling between the scalar field and the Maxwell field rather than a topological term.
Unlike previous holographic models  on commensurate lock-in states, the striped CDW in this model is generated without the need for a current density wave or magnetic field. As a result, the parity $P$ symmetry and the time-reversal $T$ symmetry are preserved.

The interplay between the striped CDW and an underlying ionic lattice has been investigated in detail and commensurate lock-in states are observed when the lattice amplitude becomes large enough. This lock-in effect forces an alignment between the period of stripes and that of lattices, leading to significant consequences on transport properties of electrons. One prominent phenomenon is the occurrence of metal-insulator transition when CDW is presented in the system. We have computed the optical conductivity as a function of the frequency for both unlock-in and lock-in states. In both cases, we have observed the  Drude peaks  as the temperature is dropped down below $T_c$.

Notably, we have also revealed two intriguing features of the commensurate lock-in states in this model. One is that its DC resistivity exhibits an algebraic increase more precisely, in agreement with previous holographic models of commensurate lock-in and experimental observations \cite{Andrade:2017ghg,Boebinger:1996,Laliberte:2016,Shi:2021}.

The present work indicates that the commensurate lock-in effect can be implemented in a general class of holographic models, and these models may exhibit abundant and distinct phenomena so as to enrich our knowledge of the interplay between the ionic lattices and electronic lattices in holographic approach. It is worthwhile to point out that there are several promising avenues for future work based on this study. The model could be made more directly applicable to real materials by incorporating impurities or disorders and studying their interplay with the lock-in effects. Also, given the intimate connection to high Tc superconductivity, studying the superconductivity phase transition based on this model would be of particular interest. Finally, we may consider the incommensurate states in this model and investigate the phase transition from a commensurate state to an incommensurate state.

\section*{Acknowledgments}
We are very grateful to Kai Li, Weijia Li, Chao Niu and Jianpin Wu for helpful discussions. This work is supported in part by the Natural Science Foundation of China under Grant No. 12035016, 12275275, 12405067 and 12475054. It is also supported by the Beijing Natural Science Foundation under Grant No. 1222031, the Innovative Projects of Science and Technology at IHEP, and by the Science and Technology Planning Project of Guangzhou (202201010655).

\appendix
 \section{Numerical techniques}

            In our investigation, we address the numerical treatment of equations in two-dimensional space $(x,z)$ plane. Here, the $x$-axis is periodic, whereas the $z$-axis extends from $0$ to $1$. Our methodology is a combination of tailored numerical methods for each axis to adapt optimally to their unique characteristics: for the $x$-axis, the Fourier series approach is employed, whereas for the $z$-axis, we rely on Chebyshev polynomials with Gauss-Lobatto points.

 \subsection{Detailed Approach}

  \begin{enumerate}
              \item {\bf Fourier Methods in $x$-Direction with Evenly Spaced Points}: The essence of employing Fourier series methods to analyze the $x$-direction stems from the periodic nature of the $x$-variable. Fourier series excel in decomposing periodic functions into a sum of sine and cosine components, each of which captures a specific frequency component of the original function. For an effective and accurate Fourier analysis, it is imperative to sample points evenly along the $x$-axis. This evenly spaced sampling ensures that each sine and cosine component is represented accurately, avoiding phenomena such as aliasing, where higher frequency components are incorrectly represented as lower frequency ones due to insufficient sampling rate.
              \item {\bf Chebyshev Methods and Gauss-Lobatto Points in $z$-Direction}: To effectively handle the $z$-axis, we incorporate Chebyshev polynomials, known for their efficacy in approximating functions over finite intervals with high accuracy. Importantly, our approach is enriched by the use of Gauss-Lobatto quadrature points. These points are carefully chosen abscissas that include the endpoints of the interval, thereby accommodating boundary conditions seamlessly. The inclusion of Gauss-Lobatto points is crucial for applications where precise integration over an interval is essential, particularly in the context of differential equations where boundary conditions play important roles.
              \item {\bf Solving Nonlinear PDEs with Newton-Raphson Iteration}: To tackle the nonlinear partial differential equations arising in the gravitational system under study, we implement the Newton-Raphson iteration method. The Newton-Raphson method is a powerful technique for finding successively better approximations to the roots (or zeroes) of functions. Applied to nonlinear PDEs, this iterative process begins with an initial guess for the solution, which is then refined incrementally to approach the true solution. Specifically, at each iteration, the method linearizes the nonlinear problem around the current guess through the calculation of the Jacobian matrix, representing the system's partial derivatives. This enables us to solve a linear approximation of the PDE for updates to the solution, systematically driving down the residuals - the differences between the left and right sides of the equations. The process is repeated until the solution converges to a satisfactory level of accuracy. To encapsulate the overall level of discrepancy across all points and equations, the norm of these residuals is computed. The norm represents a measure of the magnitude of the discrepancies, providing a single scalar value that indicates the aggregate level of error present in the numerical solution. In our paper, the residual is approximately $10^{-7}$, which signifies that, on average, the numerical solutions closely align with the expected values dictated by the EOMs within the considered computational domain. This small residual magnitude suggests that our numerical method is achieving high accuracy and convergence in capturing the underlying dynamics of the gravitational system under investigation.
  \end{enumerate}

 \subsection{Computational Details}

Our computational grid encompasses up to $20$ points along the $x$-axis and $24$ points along the $z$-axis. This selection is the culmination of a meticulous incremental refinement process, which corroborates the convergence and efficiency of our chosen grid resolution. Transitioning from coarser to denser grids, we have discerned that this grid configuration strikes an optimal balance, ensuring computational tractability while preserving the integrity of our results.

The computations are executed within the Mathematica environment, leveraging its sophisticated numerical and symbolic capabilities. Remarkably, all computational tasks are conducted on a standard laptop CPU with 32GB RAM, illustrating that our methodology does not impose prohibitive computational demands. This underscores the practical applicability of our approach, making it accessible without necessitating high-performance computing infrastructure.

In summary, our exploration validates a pragmatic and robust numerical scheme for addressing problems defined in the $(x,z)$ plane. By judiciously integrating Fourier methods with Chebyshev polynomials and Gauss-Lobatto points, we achieve a high degree of accuracy and computational efficiency. This methodology not only ensures the fidelity of our numerical solutions but also highlights the adaptability of our approach to tackle complex physical phenomena with readily available computational resources.

\begin{figure}[h]
  \centering
  \includegraphics[width=0.45\textwidth]{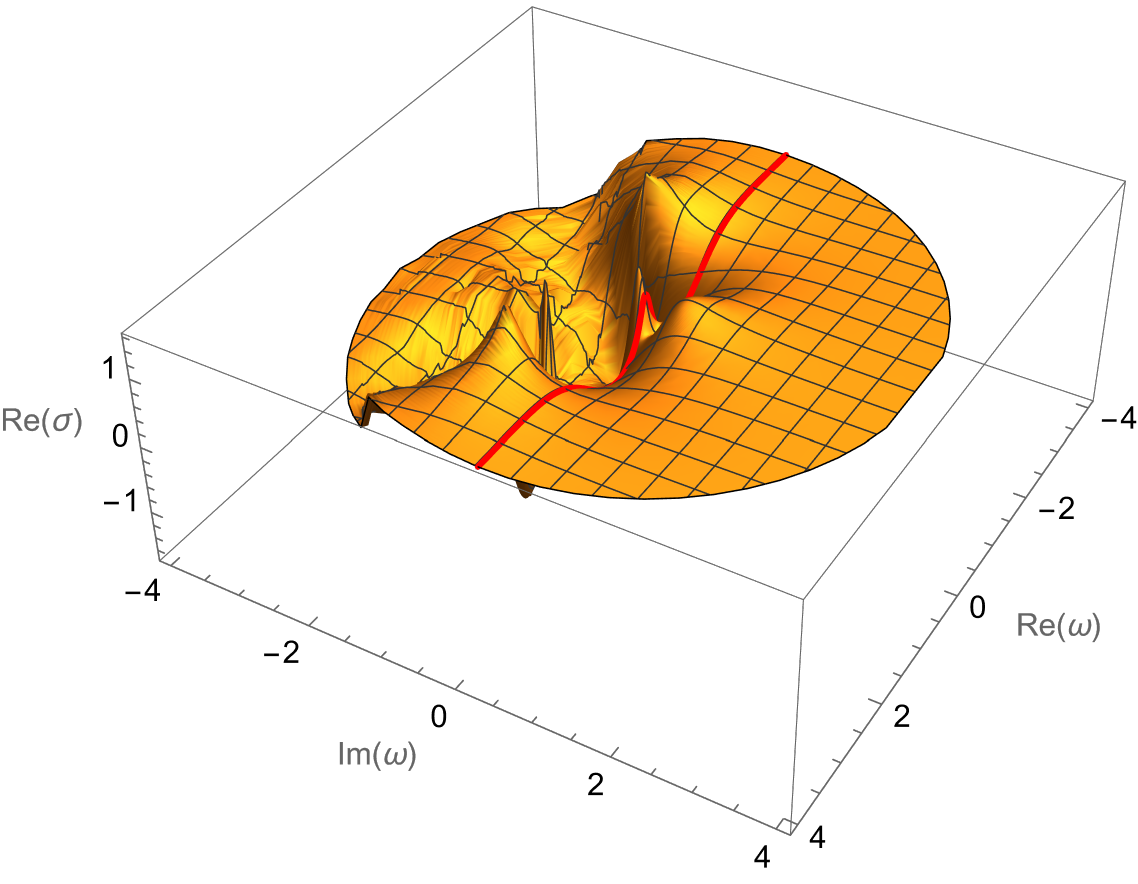}
  \includegraphics[width=0.45\textwidth]{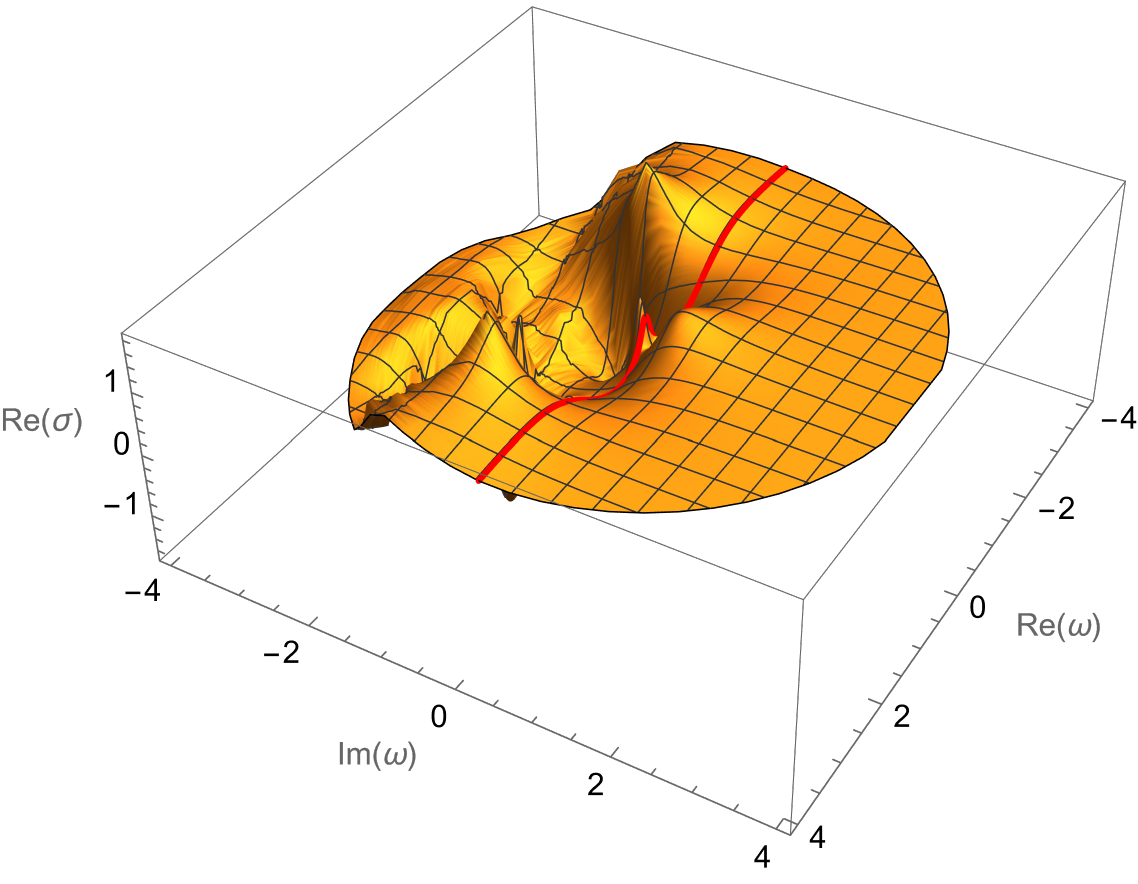}
  \caption{
    \label{fig:complex_frequency_plots}
    Complex frequency domain plots demonstrating the poles of the AC conductivity. The left plot corresponds to the system under condition $\lambda = 0.4$, and the right plot corresponds to condition $\lambda = 1.3$.
  }
\end{figure}

\section{Conductivity in the complex frequency plane}\label{app:qnm-analysis}

In this appendix, we analyze the quasinormal modes (QNMs) in detail to verify the stability of our conductivity results. We examine the analytic structure of the complex frequency plane, scanning tens of thousands of points across a broad parameter space. The findings are summarized in Fig. \ref{fig:complex_frequency_plots}, where we focus on the poles of the conductivity for two representative values of the coupling parameter: $\lambda = 0.4$ (left panel) and $\lambda = 1.3$ (right panel). In both plots, the red line represents the real frequency axis ($\mathrm{Im}[\omega] = 0$).

The dynamical stability of the system is confirmed by the absence of poles in the upper half-plane, which would correspond to exponentially growing modes. Instead, all poles are located in the lower half-plane, for both values of $\lambda$. This behavior demonstrates that the system is stable under these conditions, consistent with the stability criteria for dissipative systems.

A notable feature appeared as a bump in the upper half-plane, which could have been misinterpreted as a pole. To address this, we performed extremely dense sampling in this region to investigate its nature. The dense sampling confirmed that this feature is smooth and not indicative of a true pole. By ensuring all poles are restricted to the lower half-plane, we further verify the stability of the system and the reliability of our conductivity measurements.

The consistent location of poles exclusively in the lower half-plane across various parameter ranges reinforces our conclusion that the system remains dynamically stable.

\end{document}